\documentclass[%
 reprint,
 superscriptaddress,
 amsmath,amssymb,
 aps,longbibliography
]{revtex4-1}

\usepackage{graphicx}% Include figure files
\usepackage{dcolumn}% Align table columns on decimal point
\usepackage{bm}% bold math
\usepackage{hyperref}% add hypertext capabilities
\usepackage[mathlines]{lineno}% Enable numbering of text and display math
\usepackage[bottom]{footmisc}
\usepackage[super]{nth}%
\usepackage{color}
\usepackage{hhline}
\begin{document}
\raggedbottom

%\preprint{AIP/123-QED}
\title{Hyperuniformity and wave localization in pinwheel scattering arrays}% Force line breaks with \\
% Sgrigno
\author{F. Sgrignuoli}
\affiliation{Department of Electrical and Computer Engineering, Boston University, 8 Saint Mary\textsc{\char13}s Street, Boston, Massachusetts 02215, USA}
% Luca
\author{L. Dal Negro}
\email{dalnegro@bu.edu}
\affiliation{Department of Electrical and Computer Engineering, Boston University, 8 Saint Mary\textsc{\char13}s Street, Boston, Massachusetts 02215, USA}
\affiliation{Division of Material Science and Engineering, Boston University, 15 Saint Mary\textsc{\char13}s Street, Brookline, Massachusetts 02446, USA}
\affiliation{Department of Physics, Boston University, 590 Commonwealth Avenue, Boston, Massachusetts 02215, USA}
%%%%%%%%%%%%%%%%%%%%%%%%%%%%%%%%%%%%%%%%%%%%%%%%%%%%%%%%%%%%%%%%%%%%%
%%%%%%%%%%%%%%%%%%%%%%%                  Abstract                      %%%%%%%%%%%%%%%%%%%%%%%%%%%%%
%%%%%%%%%%%%%%%%%%%%%%%%%%%%%%%%%%%%%%%%%%%%%%%%%%%%%%%%%%%%%%%%%%%%%
\begin{abstract}
We investigate the structural and spectral properties of deterministic aperiodic arrays designed from the statistically isotropic pinwheel tiling. By studying the scaling of  the cumulative integral of its structure factor in combination with higher-order structural correlation analysis we conclude that pinwheel arrays belong to the weakly hyperuniformity class. Moreover, by solving the multiple scattering problem for electric point dipoles using the rigorous Green\textsc{\char13}s matrix theory, we demonstrate a clear transition from diffusive transport to localization behavior. This is shown by studying the Thouless number as a function of the scattering strength and the spectral statistics of the scattering resonances. Surprisingly, despite the absence of sharp diffraction peaks, clear spectral gaps are discovered in the density of states of pinwheel arrays that manifest a distinctive long-range order. Furthermore, the level spacing statistics at large optical density exhibits a sharp transition from level repulsion to the Poisson behavior, consistently with the onset of the wave localization regime. 
Our findings reveal the importance of hyperuniform aperiodic structures with statistically isotropic $k$-space for the engineering of enhanced light-matter interaction and localization properties.
\end{abstract}

\pacs{Valid PACS appear here}
\keywords{Suggested keywords}
%Use showkeys class option if keyword
\maketitle
\section{Introduction}
% ==================================    Introduction    ==========================================================
%%%%%%%%%%%%%%%%%%%%%%%%%%%%%%%%%%%%%%%%%%%%%%%%%%%%%%%%%%%%%%%%%%%%%%%%%%%%%
%%%%%%%%%%%%%%%%%%%%%%%%%%%%%%%%%%%%%%%%%%%%%%%%%%%%%%%%%%%%%%%%%%%%%%%%%%%%%
Isotropic nanostructures with circular symmetry in $k$-space (i.e., k-space isotropy) have been proposed to achieve more robust optical bandgaps \cite{Florescu,Pollard,Froufe}, enhance the efficiency and directionality of LEDs \cite{Wiesmann,Wierer,LawrenceJAP,Gorsky1,Gorsky} and optical lasers \cite{Notomi}, create angle-insensitive structural coloration \cite{Lee,Boriskina}, and novel cavity-enhanced platforms for quantum photonics \cite{Trojak,Trojak2}. Isotropy in the $k$-space occurs in disordered systems with long-range correlations (i.e., stealthy hyperuniform \cite{Florescu,TorquatoReview} and optimized isotropic scattering arrays \cite{Gorsky}), quasi-periodic structures \cite{Notomi,David,Hagelstein}, and aperiodic media beyond quasicrystals that feature nearly continuous circular $k$-space symmetry, such as Vogel spirals \cite{Guo,Trevino,DalNegroVogel,Lawrence} and the pinwheel tiling \cite{Pierro,Lee,LawrenceJAP}.  

In this work, we study the transport and localization properties of optical waves in aperiodic arrays of scattering dipoles positioned at the vertices of the pinwheel tiling \cite{Radin1994,RadinBook}. A two-dimensional (2D) pinwheel tiling is an aperiodic tiling of the plane constructed by a deterministic inflation rule that produces rotated copies of a triangular prototile in infinitely many distinct orientations, giving rise to  isotropic $k$-space (see Section \ref{Sec1} for more details). Motivated by the recent discovery that two-dimensional (2D) disordered  hyperuniform and isotropic scattering media support a localization transition for transverse magnetic (TM) waves \cite{Aubry,FroufePNAS}, we ask whether wave localization can also be achieved in deterministic and isotropic pinwheel arrays. To investigate wave localization in an open 2D scattering environment ( i.e., a system with in-plane radiation losses), we applied the rigorous Green\textsc{\char13}s matrix spectral method that enables a systematic investigation of complex scattering resonances and their spectral statistics \cite{Lagendijk,RusekPRE2D}. Specifically, by studying the Thouless number $g$ and the first-neighbor level-spacing statistics for different values of the optical density our work demonstrates a transition from the diffusive to the localized transport regime accompanied by a crossover from level repulsion to level clustering behavior. Moreover, spectral gaps are discovered at large optical density by studying the density of optical states (DOS) of the pinwheel structure and the formation of localized resonances is observed around the band-edges, similarly to disordered band-gap materials \cite{Aubry,FroufePNAS,John,John2,Skipetrov2020}. 

\section{Geometrical properties of Pinwheel arrays}\label{Sec1}
% ==================================  Fig.1 discussion   ========================================================== 
The aperiodic pinwheel tiling is a hierarchical structure iteratively generated from a simple inflation rule that decomposes a rectangular triangle with edge lengths proportional to 1, 2, and $\sqrt{5}$ (i.e., the prototile) into five congruent copies \cite{Baake,Senechal,Radin1994,RadinBook}. An infinite tiling is produced by iterating and rescaling this linear decomposition. Since the inflation rule also involves a rotation by an angle that is an irrational modulo $2\pi$, the resulting tiling contains copies of the original prototile arranged in infinitely many distinct orientations resulting in statistical isotropy, referred to as the ``pinwheel phenomenon"  \cite{Baake,Senechal}. As a consequence, the pinwheel tiling displays an infinity-fold rotational symmetry. The pinwheel tiling, first introduced by Conway and Radin, does not contain  discrete components in its diffraction spectrum, which is conjectured to be absolutely continuous although it is presently unknown whether there exists a singular-continuous component as well \cite{Radin1994,RadinBook,Senechal}. 
%Moreover, the diffraction spectrum of pinwheel has no discrete components \cite{Radin1994}, and it is presently unknown if it is continuous \cite{Senechal}.

\begin{figure}[t!]
\centering
\includegraphics[width=\columnwidth]{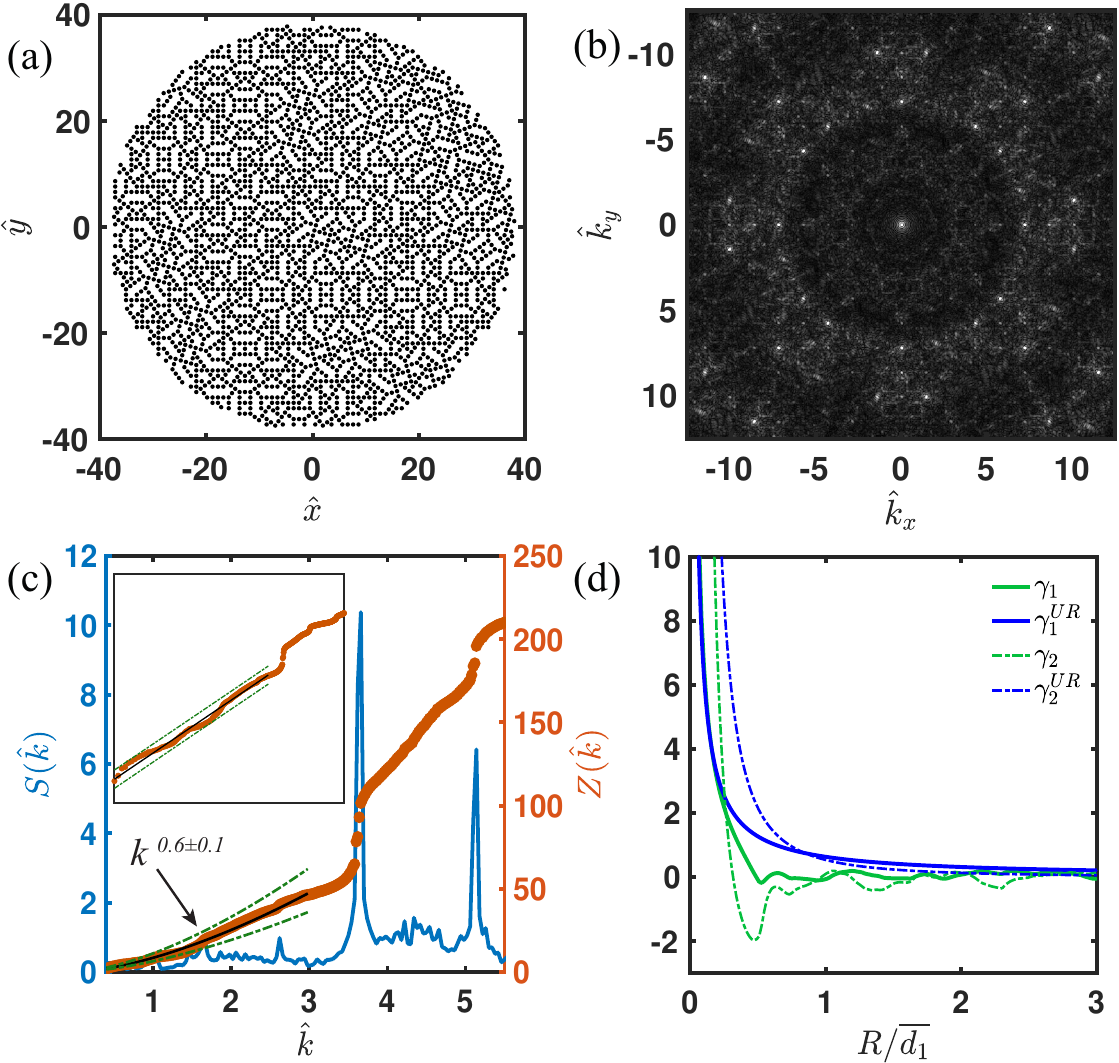}
\caption{Panel (a) shows 3907 scatters spatially arranged in a pinwheel geometry. $\hat{x}$ and $\hat{y}$ indicate normalized coordinates with respect to the average inter-particle distance $d_1$. Panel (b)  presents the structure factor of the array shown in (a) [cubic root is taken to enhance contrast]. Panel (c) displays the azimuthally averaged structure factor (left blue axis) and cumulative intensity function (right orange axis). The inset reports a zoom-in-view of the scaling of the $Z(\hat{k})$. Here, $\hat{k}$ indicates the product of the wavenumber $k$ with $d_1$. The black line is the power-law fit within the range $0.1<\hat{k}<3$, while the green dashed-lines represent the $95\%$ prediction interval. Panel (d) compares the $\gamma_1$ and $\gamma_2$ functions of the pinwheel array (green lines) to the analytical trends of uncorrelated Poisson processes (blue curves) \cite{Mehta,Sgrignuoli_PRB2020}.}
\label{Fig1}
\end{figure}

The pinwheel array shown in Fig.\ref{Fig1}\,(a) is obtained by positioning one scattering electric dipole at each node of the corresponding tiling. The resulting scattering array is aperiodic and features a diffraction pattern, proportional to the structure factor $S(\textbf{k})$ shown in Fig.\ref{Fig1}\,(b), that is  essentially continuous except from few bright spots that are due to the finite size of the system \cite{Senechal,Lee}. In the infinite size limit, the tiles occur in infinitely many orientations and the rotational symmetry of the spectrum becomes continuous \cite{Senechal}. In fact, Radin has mathematically shown that there are no discrete components in the diffraction spectrum of the pinwheel tiling \cite{Radin1994}, but we remark again that it is unknown whether there is also a singular continuous component \cite{Senechal}. Generally, the structure factor displays a highly-structured diffuse background that manifests $k$-space isotropy, i.e., a circularly symmetric diffuse scattering background that resembles a powder diffraction pattern \cite{Moody,BaakePinwh}.

Since the structure factor of the pinwheel array is isotropic, we show in Fig.\ref{Fig1}\,(c) its azimuthal average computed  according to the formula \cite{Gorsky}:
\begin{equation}
S_\theta(k)=\frac{1}{2\pi}\int_0^{2\pi} S(k,\theta)d\theta
\end{equation} 
The sharp peaks in $S_\theta(k)$, shown by the blue line, quantify the overall isotropic scattering strength of the pinwheel array that has been recently exploited in engineering applications to radiation extraction \cite{LawrenceJAP,Pierro} and bright structural coloration of metal surfaces \cite{Lee}.
Additionally, the point patterns obtained from the pinwheel tiling are known to be hyperuniform \cite{Gabrielli,Torquato}. 
Hyperuniformity is a correlated state of matter characterized by the suppression of long-wavelength density fluctuations \cite{Torquato,TorquatoReview}. Hyperuniformity has been recently shown to play an important role in light localization phenomena for both disordered and deterministic systems \cite{Sgrignuoli_PRB2020,FroufePNAS,Aubry,Haberko,Yu}. 

Hyperuniform systems can be classified according to three main categories depending on the power-law scaling of their structure factors in the vicinity of the $k$-space origin, i.e.,  $S(\bold{k})\sim|\bold{k}|^\alpha$ in the limit $\bold{k}\rightarrow0$ \cite{TorquatoReview}. Specifically, $\alpha>1$, $\alpha=1$, and $0<\alpha<1$ define, respectively,  the strong (Class I), the logarithmic  (Class II), and the weak (Class III) hyperuniform class \cite{TorquatoReview}.
Currently, the hyperuniformity class of the pinwheel arrays is not known.  

In order to better understand the type of hyperuniformity that characterizes pinwheel arrays we have investigated the scaling behavior of the cumulative diffraction power $Z(k)$ that can be directly computed from the structure factor as follows \cite{Ouguz}:
\begin{equation}\label{Zk}
Z(k)=\int_0^{k}\int_0^{2\pi}S(k^\prime,\theta) d\theta dk^\prime
\end{equation} 
Equation\,(\ref{Zk}) scales as $k^{\alpha+1}$ when $k\rightarrow0$ if the array is hyperuniform. The $\alpha$ coefficient can then be estimated from a linear-fit in double logarithmic scale \cite{Ouguz}, thus identifying the corresponding hyperuniformity class of the investigated structure. The orange lines in Fig.\,\ref{Fig1}\,(c) and in the corresponding inset show the scaling behavior of the cumulative integral of the structure factor. The black lines is the result of the power-law fit and the green dashed-lines represent the 95$\%$ prediction interval. Our results demonstrate that the pinwheel arrays are weakly hyperuniform structures characterized by $\alpha=0.6\pm0.1$. 

General hyperuniform structures must obey the following direct-space sum rule \cite{LucaBibbiaBook}:
\begin{equation}\label{DSrule}
\rho\int_{\mathbb{R}^d}h(\textbf{r})d\textbf{r}=-1
\end{equation}
where $\rho$ is the number density, while $h(\textbf{r})$ is the total correlation function that vanishes in the absence of spatial correlations in the system \cite{Torquato}. Equation\,(\ref{DSrule}) implies that in hyperuniform systems the total correlation function must become negative for some values of $\textbf{r}$ \cite{LucaBibbiaBook}. A general approach to identify regions of negative structural correlations, developed initially to investigate correlations in the energy levels of nuclear spectra \cite{Mehta,Bohigas}, is based on the analysis of skewness (i.e., $\gamma_1$) and excess kurtosis (i.e., $\gamma_2$) functions \cite{Torquato2020local,Sgrignuoli_PRB2020}. These statistical quantities are defined in terms of the moments: 
\begin{equation}\label{moments}
\mu_j=\Big\langle \Big(n-\langle n\rangle\Big)^j\Big\rangle
\end{equation}
where $n$ is the number of elements in an interval of length $L$ and $\langle\cdots\rangle$ represents an average taken over many such intervals throughout the entire system \cite{Sgrignuoli_PRB2020}. Besides providing a precise characterization of level repulsion and long-range order, $\gamma_1$ and $\gamma_2$ are sensitive to three-level $\mu_3$ and four-level $\mu_4$ structural correlations. In fact, $\gamma_1$ and $\gamma_2$ are equal to $\mu_3\mu_2^{-3/2}$ and $\mu_4\mu_2^{-2}-3$, respectively \cite{Mehta}. 

In Fig\,\ref{Fig1}\,(d), we compare the size-scaling behavior of the $\gamma_1$ and $\gamma_2$ functions of the pinwheel array (green curves) with the analytical expressions  (blue curves) corresponding to a uniform random (UR) point pattern, which can be expressed as $\gamma_1^{UR}=(R/d_1)^{-1}/2\sqrt{\rho}$ and $\gamma_2^{UR}=(R/d_1)^{-2}/3\rho$ \cite{Mehta,Sgrignuoli_PRB2020}. We observe that the pinwheel array exhibits a range where $\gamma_2$ is oscillatory and negative, indicating the presence of strong structural correlations with repulsion behavior \cite{Torquato2020local,Sgrignuoli_PRB2020}. On the contrary, UR systems do not feature structural correlations up to fourth-order correlation functions \cite{Mehta}. 
Our scaling analysis further demonstrates the hyperuniform nature of the investigated pinwheel arrays and unveils a prominent anti-clustering behavior.

\section{Spectral properties of pinwheel arrays}
We now investigate the wave transport and localization properties of TM-polarized electric dipoles that are spatially arranged as in Fig.\ref{Fig1}\,(a). Multiple scattering effects in two spatial dimensions (i.e., for cylindrical waves) are studied by analyzing the spectral properties of the Green\textsc{\char13}s matrix defined as:
%%%%%%%%%%%%%%%%%%%%%% Green Matrix %%%%%%%%%%%%%%%%%%%%%% 
\begin{equation}\label{Green}
G_{ij}=i\left(\delta_{ij}+\tilde{G}_{ij}\right)
\end{equation}
%%%%%%%%%%%%%%%%%%%%%%%%%%%%%%%%%%%%%%%%%%%%%%%%%%%
where the elements $\tilde{G}_{ij}$ are given by \cite{RusekPRE2D}:
%%%%%%%%%%%%%%%%%%%%%% Green Our %%%%%%%%%%%%%%%%%%%%%%%
\begin{eqnarray}\label{GreenOur}
\begin{aligned}
\tilde{G}_{ij}=\frac{2}{i\pi}K_0(-ik_0|\textbf{r}_i-\textbf{r}_j|)
\end{aligned}
\end{eqnarray}
%%%%%%%%%%%%%%%%%%%%%%%%%%%%%%%%%%%%%%%%%%%% %%%%%%
and $K_0(-ik_0|\textbf{r}_i-\textbf{r}_j|)$ denotes the modified Bessel function of the second kind, $k_0$ is the wavevector of light, and $\textbf{r}_i$ specifies the position of the $i$-th scattering dipole in the array. The non-Hermitian matrix (\ref{Green}) describes the electromagnetic coupling among the scatterers and the real and imaginary part of its complex eigenvalues $\Lambda_n$ ($n\in$ 1, 2, $\cdots$N) correspond to the detuned frequency $(\omega_0-\omega)$ and  decay rate $\Gamma_n$ (both normalized to the resonant width $\Gamma_0$ of an isolated dipole) of the scattering resonances of the system \cite{RusekPRE2D,Lagendijk}. This formalism accounts for all the multiple scattering orders and enables the systematic study of the scattering properties of 2D waves with an electric field parallel to the invariance axis of the scatterers \cite{Leseur}. Even though the 2D model in (\ref{Green}) does not take into account the vector nature of light \cite{DalNegroElliptic,Sgrignuoli2019,Sgrignuoli2019compact,SkipetrovPRL}, it still provides useful information on light localization in 2D disorder media \cite{RusekPRE2D}, transparency in high-density hyperuniform materials \cite{Leseur}, and correctly describes the coupling between one or several quantum emitters embedded in structured dielectric environments  \cite{Caze,Bouchet}. 

\begin{figure}[b!]
\centering
\includegraphics[width=\columnwidth]{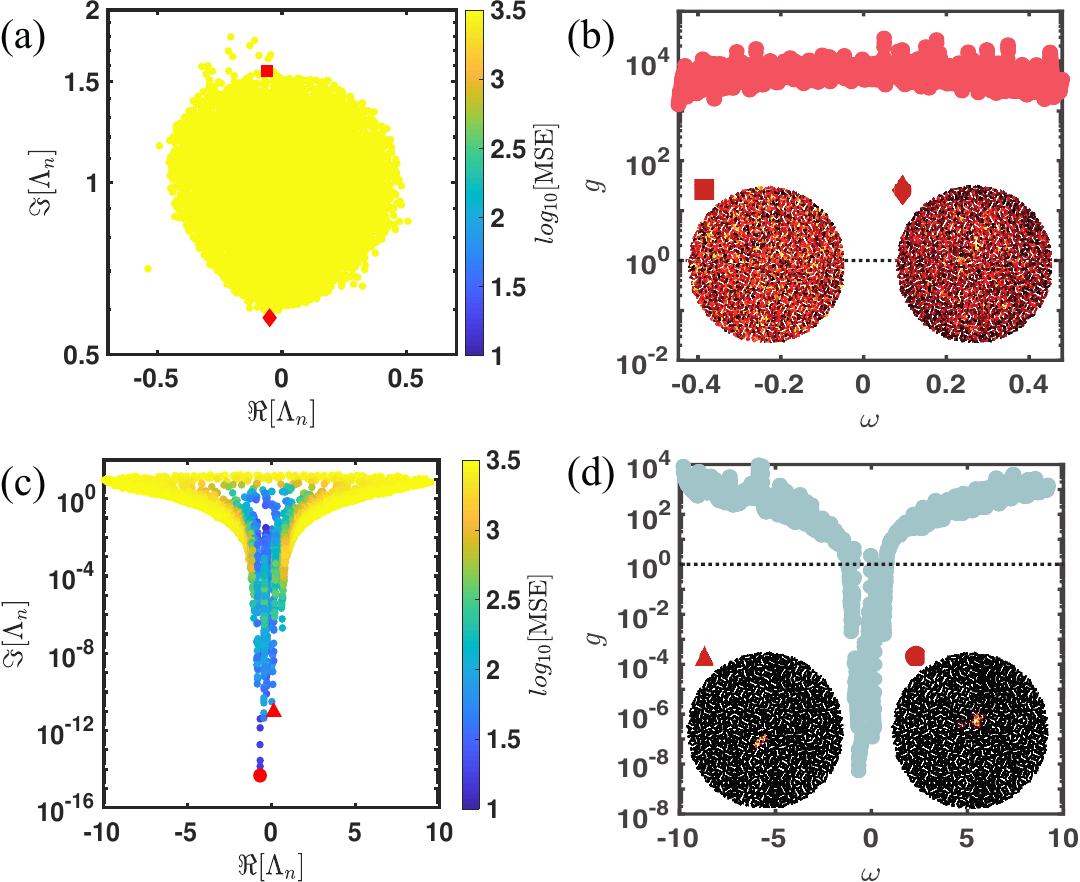}
\caption{Panel (a) and (b) show the complex eigenvalue distribution and the Thouless number $g$ as a function of the frequency $\omega$ when $\rho\lambda^2=10^{-4}$. Panels (c) and (d) display the complex eigenvalue distribution and the Thouless number $g$ when $\rho\lambda^2$ is equal to 5. The dashed-black lines in the panel (b) and (d) identify the threshold of the diffusion-localization transition $g=1$. The different markers in the panel (a) and (c) identify the spectral positions of the representative scattering resonances reported in panel (b) and (d).} 
\label{Fig2}
\end{figure}

% ==================================  Fig.2 discussion   ==========================================================
To investigate the nature of the spectral properties of the pinwheel arrays, we analyze the distributions of the complex eigenvalues that describe the scattering resonances, the Thouless number, the level spacing statistics, and the density of optical states (DOS) for different values of the scattering strength of the system, which is quantified by its optical density $\rho\lambda^2$. Here, $\rho$ denotes the number of scatterers per the unit area, and $\lambda$ is the optical wavelength. The spectral information is derived by numerically diagonalizing the $N\times N$ Green\textsc{\char13}s matrix (\ref{GreenOur}). Figure\,(\ref{Fig2}) shows the results of this analysis at both small (i.e., $\rho\lambda^2=10^{-4}$) and large (i.e., $\rho\lambda^2=15$) optical density. In particular, Fig.\,(\ref{Fig2}) (a) and (c) display the distributions of the complex eigenvalues of the Green\textsc{\char13}s matrix (\ref{Green}) in the small and large optical density regimes, respectively. The complex eigenvalues are color-coded according to the $\log_{10}$ values of the modal spatial extent (MSE) of the corresponding eigenvectors. The MSE parameter quantifies the spatial extension of a given scattering resonance of the system \cite{SgrignuoliACS,Mirlin}. Figure\,(\ref{Fig2}) (b) and (c) display the Thouless number $g$ as a function of the frequency $\omega$ corresponding to the two scattering regimes.

To evaluate the Thouless number $g$ as a function of $\omega$, we have used the following definition \cite{Sgrignuoli2019,Sgrignuoli_PRB2020}:
\begin{equation}\label{Thouless}
g(\omega)=\frac{\overline{\delta\omega}}{\overline{\Delta\omega}}=\frac{(\overline{1/\Im[\Lambda_n]})^{-1}}{\overline{\Re[\Lambda_n]-\Re[\Lambda_{n-1}]}}
\end{equation}
We have sampled the real parts of the eigenvalues of the Green\textsc{\char13}s matrix in 500 equi-spaced intervals and we computed eq.\,(\ref{Thouless}) in each frequency sub-interval. The symbol $\overline{\{\cdots\}}$ in eq.\,(\ref{Thouless}) denotes the sub-interval averaging operation, while $\omega$ indicates the central frequency of each sub-interval. We have verified that the utilized frequency sampling resolution does not affect the presented results. 

In the low optical density regime, the complex eigenvalue distribution does not show the presence of any long-lived scattering resonance with $\Gamma_n\ll\Gamma_0$, as visible in Fig.\,\ref{Fig2}\,(a). Consistently, the corresponding Green\textsc{\char13}s matrix eigenvectors are spatially delocalized across the array. Two representative delocalized modes are shown in Fig.\,\ref{Fig2}\,(b). These resonances are labelled by the square and rhombus red markers that indicate their positions in the spectrum. Moreover, the Thouless number $g$ remains larger than unity independently of the frequency $\omega$, which indicates diffusive transport.

On the other hand, at large optical density spatially confined long-lived scattering resonances appear in the spectrum when  $\omega_n\approx\omega_0$. 
Two representative localized modes are shown in Fig.\ref{Fig2}\,(d) labelled by the circle and triangle red markers indicating their positions in the spectrum. Moreover, two dispersion branches populated by scattering resonances localized over small clusters of dipoles near the center of the array [see Fig.\ref{Fig2}\,(c)], appear in the distribution of complex eigenvalues. Figure\,\ref{Fig2}\,(c) also shows the formation of a spectral gap region where the critical scattering resonances reside. These  are spatially extended and long-lived resonances with strong spatial fluctuations at multiple length scales characterized by a power-law scaling behavior \cite{Sgrignuoli_MF,RyuPRB,macia1999,Trojak}. The formation of a spectral gap region at large optical density reflects the long-range correlated nature of the pinwheel array. Furthermore, at large optical density, we find that $g$ becomes lower than one for $\omega\approx{0}$, indicating the onset of light localization, as demonstrated in Fig.\,\ref{Fig2}\,(d). We remark that the long-lived scattering resonances that are spatially confined over few scatterers appear at the frequency positions where the Thouless number becomes lower than one. 

\begin{figure}[b!]
\centering
\includegraphics[width=\columnwidth]{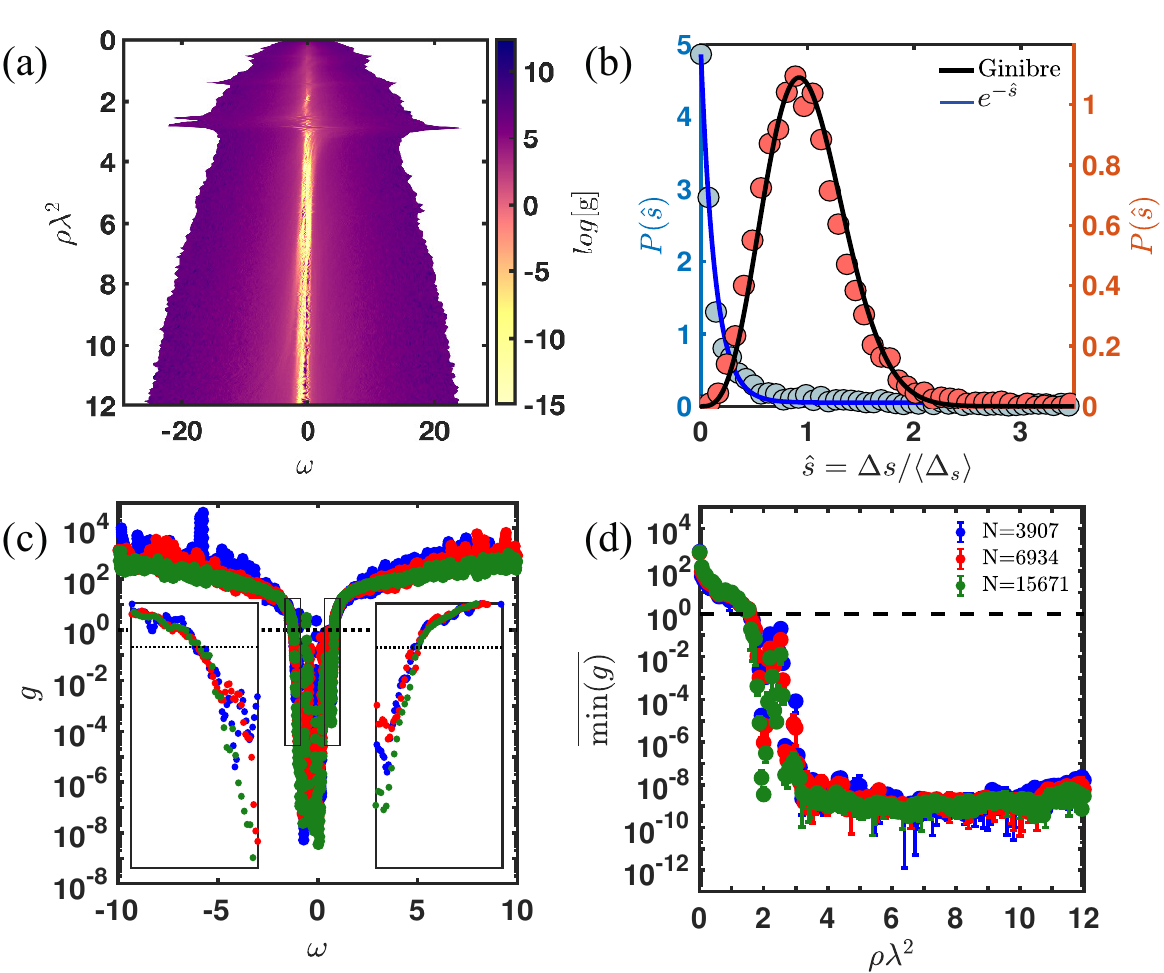}
\caption{Panel (a) shows a high resolved map of the logarithmic values of the averaged Thouless number for different $\rho\lambda^2$ values as a function of $\omega$. This map is evaluated in the range $\rho\lambda^2=[10^{-4},12]$ with a resolution of 0.06. Panel (b) shows the crossover from level repulsion to level clustering of $P(\hat{s})$. Specifically, $P(\hat{s})$ changes from the Ginibre\textsc{\char13}s statistic (black line) to the Poisson distribution (blue line) by increasing $\rho\lambda^2$.
Panel (c) reports the Thouless number as a function of the frequency $\omega$ at a given optical density $\rho\lambda^2=5$ for different system sizes. Blue, red, and green circle markers indicate a total number of scatterers of 3907, 6934, and 15671, respectively. Panels (d) displays the scaling of the minimum value of the Thouless conductance as a function of $\rho\lambda^2$.}
\label{Fig3}
\end{figure}

% ==================================  Fig.3 discussion   ==========================================================
To obtain additional insights on this localization transition, we analyze the Thouless number as a function of the normalized frequency $\omega$ for different optical density values $\rho\lambda^2$, starting from $10^{-4}$ up to 12 with a resolution of 0.06. Figure\,\ref{Fig3}\,(a) shows a high-resolution map that is color-coded according to the quantity $\ln[g]=\ln[g(\omega,\rho\lambda^2)]$. Localization phenomena begin to occur when $\rho\lambda^2\approx{1.5}$. Moreover, Fig.\,\ref{Fig3}\,(a) displays a clear dispersion branch followed by the localized scattering resonances (i.e., the yellow stripes) that are similar to the representative example previously shown in Fig.\,\ref{Fig2}\,(d). 

The transition from diffusion to localization is confirmed by the switching from level repulsion to level clustering of the quantity $P(\hat{s})$ as a function of $\rho\lambda^2$, which is demonstrated in Fig.\,\ref{Fig3}\,(b). Here, $P(\hat{s})$ denotes the probability density function of the first-neighbor level spacing distribution of the complex eigenvalues of the Green\textsc{\char13}s matrix \cite{Escalante}. It is well-established that the suppression of the level repulsion (i.e., $P(\hat{s})\rightarrow$0 when $\hat{s}$ goes to zero) indicates the transition into the localization regime for both scalar and vector waves in two-dimensional and three-dimensional disordered systems \cite{Skipetrov2015,Escalante,Mondal} as well as non-uniform aperiodic deterministic structures \cite{Sgrignuoli2019,Wang_Prime,DalNegroElliptic,Sgrignuoli_PRB2020}. When $\rho\lambda^2=10^{-4}$, the $P(\hat{s})$ of the pinwheel array, shown with red-circle markers in Fig.\,\ref{Fig3}\,(b), can be modeled by the Ginibre distribution, defined as \cite{Haake}:
\begin{equation}\label{Ginibre}
P(\hat{s})=\frac{3^4\pi^2}{2^7}s^3 \exp\left(-\frac{3^2\pi}{2^4}s^2\right)
\end{equation}
We emphasize that the black curve in Fig.\,\ref{Fig3}\,(b) does not result from data fitting but is obtained using directly eq.\,(\ref{Ginibre}). The Ginibre model extends the analysis of level repulsion to non-Hermitian random matrices \cite{Haake}, which correspond to open-scattering systems. Our analysis based on the  Ginibre distribution demonstrates that the level spacing of the  pinwheel\textsc{\char13}s complex eigenvalues exhibits cubic level repulsion in the low scattering regime. On the other hand, at large optical density (i.e., $\rho\lambda^2=5$), the distribution of the level spacing statistics changes drastically, showing level clustering. This is demonstrated in Fig.\,\ref{Fig3}\,(b) by the blue-circle markers. These data are well-described by the Poisson distribution $e^{-\hat{s}}$, which is typically associated to non-interacting, exponentially localized energy levels \cite{Haake,Mehta}.

Figure\,\ref{Fig3} (c) displays the Thouless number as a function of the frequency $\omega$ in the localized regime for pinwheel systems with an increasing number $N$ of scatterers (i.e., for an increasing size). The blue, red, and green markers refer to arrays with 3907, 6934, and 15671 particles, respectively. All these curves cross at the threshold value $g=1$ at two points, independently from $N$, as displayed more clearly in the insets. The abscissas of these two points can be taken as a rough estimate of the two mobility edges that characterize the onset of the localization transition \cite{SkipetrovPRL}. Interestingly, the same behavior was recently observed in the propagation of scalar waves through a three-dimensional ensemble of resonant point scatterers \cite{SkipetrovPRB}. Figure\,\ref{Fig3} (d) shows the behavior of the minimum value of the Thouless number as a function of  $\rho\lambda^2$ for three different values of $N$. Specifically, we have evaluated $g=g(\omega)$ by using eq.\,(\ref{Thouless}) for each $\rho\lambda^2$ value and we have repeated this procedure for different frequency resolutions used in the Thouless number computation. The circle markers and the error bars in Fig.\,\ref{Fig3}\,(d) are the averaged values and the standard deviations corresponding to different frequency resolutions. The scaling of $\overline{\min[g]}$ as a function of the optical density switches from $\overline{\min[g]}>1$ to $\overline{\min[g]}<1$, demonstrating the diffusion to localization transition. 

Even though the pinwheel array manifests a characteristic hyperuniform long-range order, the discovered wave localization transition shares similar properties with the Anderson transition in disordered media. In particular, in 2D random media the Thouless number drops below unity and the probability density of the level spacing switches from the Ginibre distribution, describing level repulsion in the diffusive regime, to the Poisson distribution, which is characteristic of level clustering in the localization regime. 

A simple justification of the observed localization threshold can be obtained by estimating the localization length $\xi$. For a uniform and isotropic random system, the characteristic localization length is predicted to be \cite{Sheng,Gupta}: 
\begin{equation}\label{xi}
\xi\sim l_t\exp[\pi \Re(k_e)l_t/2]
\end{equation}
with $l_t$ the transport mean free-path and $\Re(k_e)$ the real part of the effective wavenumber in the medium. Although the numerical factor in eq.\,(\ref{xi}) may not be accurate \cite{Sheng,Gupta}, it nevertheless tells us that the localization length in 2D systems is an exponential function of $l_t$ and can be extremely large in the weak scattering regime (i.e., in the low optical density regime). Moreover, for isotropic scattering systems like the ones considered in this work, the transport mean free path coincides with the scattering mean free path $l_s$, i.e., $l_t=l_s=1/\rho\sigma_d$. Here, $\sigma_d$ is the cross-section of a single point scatterer, which is related to the 2D electric polarizability $\alpha(\omega)$ \cite{Caze}. At resonance, $\sigma_d$ is equal to $k_0^3|\alpha(\omega_0)|^2/4$ \cite{Lagendijk,Leseur}. Considering that, under the effective medium theory, $k_e$ can be approximated as $k_0+i/(2l_s)$ \cite{Leseur,Caze}, the eq.\,(\ref{xi}) can be rewritten as $\pi\lambda\exp[\pi^3/(2\rho\lambda^2)]/(2\rho\lambda^2)$, which relates the localization length of isotropic structures with their optical density. 
In order to simply account for the discovered transition we have to consider the ratio of $\xi/{L}$ where $L$ is the linear size of the system. We immediately realize that when $\rho\lambda^2=10^{-4}$, $L/\xi\gg1$ and the transport regime is diffusive, while for $\rho\lambda^2=5$, $\xi/L=0.2$ that is consistent with the onset of the localization regime.

\begin{figure}[b!]
\centering
\includegraphics[width=\columnwidth]{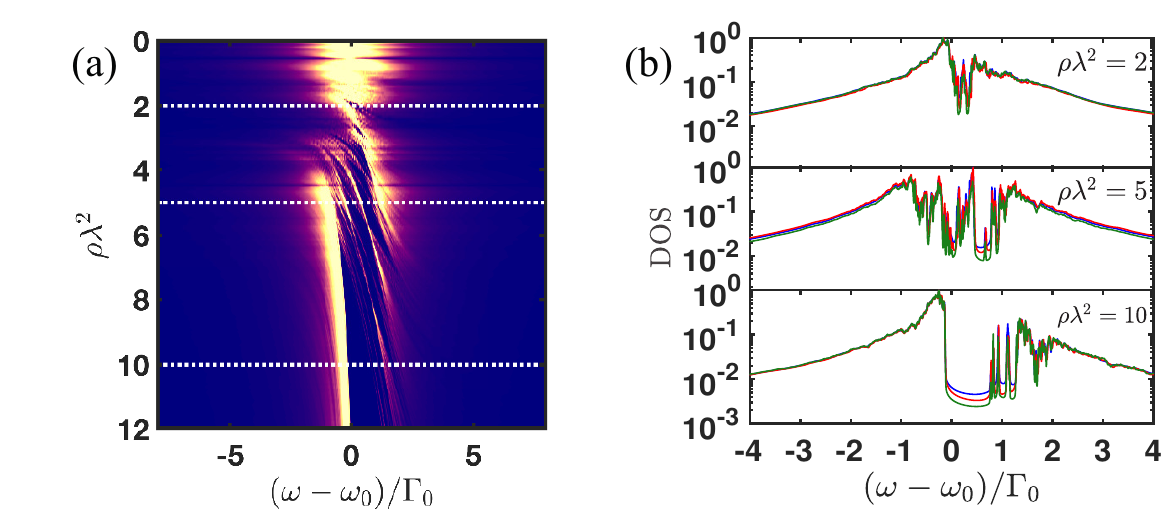}
\caption{Panel (a) displays a highly resolved map of the DOS for as a function of the normalized frequency $(\omega-\omega_0)/\Gamma_0$ and for different $\rho\lambda^2$ values. Panel (b) shows the scaling of the DOS with respect to $N$  for three selected optical density values, marked with white dashed-lines in panel (a). We have considered 3907 (blue lines), 6934 (red lines), and 15671 (green lines) scatterers, respectively.}
\label{Fig4}
\end{figure}

% ==================================  Fig.4 discussion   ==========================================================
To further understand the physical mechanism beyond the localization transition we study the behavior of the density of states (DOS) associated to the spectral distribution of scattering resonances. To evaluate the DOS within the Green\textsc{\char13}s matrix spectral method we used the approach introduced in refs.\cite{Skipetrov2020,Skipetrov2020finite} for the scalar case. In particular, the DOS can be rigorously obtained from the knowledge of the complex eigenvalues $\Lambda_n$ of the Green\textsc{\char13}s matrix according to:
\begin{equation}\label{DOS}
DOS=\frac{1}{N\pi}\sum_{n=1}^{N}\frac{\Gamma_n/2}{(\omega+\omega_n)^2+(\Gamma_m/2)^2}
\end{equation}
where $N$ is the number of scatterers in the system, $\omega_m=\omega_0-\Gamma_0\Re[\Lambda_n]/2$, and $\Gamma_m=\Gamma_0\Im[\Lambda_n]$ \cite{Skipetrov2020,Skipetrov2020finite}. In Figures\,\ref{Fig4}\,(a) and (b) we show the behavior of the DOS as a function of frequency $(\omega-\omega_0)/\Gamma_0$ for different values of the optical density $\rho\lambda^2$. As discussed in detail in ref.\cite{Skipetrov2020finite}, eq.\,(\ref{DOS}) considers only the atomic component in the excitations of the coupled system atoms+light. Therefore, eq.\,(\ref{DOS}) does not converge to the DOS of the free electromagnetic field at low optical density. Instead, it approaches a simple Lorentzian function centered at $\omega=\omega_0$. This behavior is shown in Fig.\,\ref{Fig4}\,(a) up to $\rho\lambda^2$ equal to 1.5, where we plot a high-resolution map color-coded according to eq.(\ref{DOS}) as a function of $(\omega-\omega_0)/\Gamma_0$ and $\rho\lambda^2$. 
At optical densities  larger than the threshold value $\rho\lambda^2=1.5$, our results show the formation of local band gap regions populated by band-edge localized modes.  Moreover, we notice that the frequency positions of the scattering resonances that minimize the $g$ values [see Fig.\,\ref{Fig3}\,(a)] correspond to a spectral region where the DOS is relatively low. This shows  that, analogously to random systems \cite{Aubry,FroufePNAS,John,John2,Skipetrov2020}, wave localization in the pinwheel array is enhanced around the spectral regions of low DOS, i.e., close to the pseudo-band gaps of the system.
Finally, we investigate the DOS for different system sizes. Specifically, Fig.\,\ref{Fig4}\,(b) shows the DOS behavior as a function of $(\omega-\omega_0)/\Gamma_0$ when the number of scatterers $N$ is equal to 3907 (blue line), 6934 (red line), and 15671 (olive-green line), respectively. It is well-known that the vanishing of the DOS in a gap only occurs in the infinite-size limit \cite{Joannopoulos}. In our case, the DOS inside the gap is different from zero due to the finite size of the investigated systems \cite{hasan2018finite}. This fact is visible in Fig.\,\ref{Fig4}\,(b) for $\rho\lambda^2=10$ where an increase of the size of the system indeed reduces the DOS value inside the gap region. Our findings demonstrate that TM-polarized electric dipoles arranged in a pinwheel array support a transition from wave diffusion to localization that occurs due to the suppression of the DOS near spectral band-edge regions. 
 
% =====================================   Conclusions   =========================================================
\section{Conclusions}
In conclusion, we have systematically investigated the structural and spectral properties of statistically isotropic pinwheel arrays that we found to be weakly hyperuniform systems.  Moreover, we have unveiled a transition, similar to the Anderson one, from a diffusive to a localized regime by evaluating the Thouless number $g$ and studying the first-neighbor level-spacing statistics of the complex eigenvalues of the Green\textsc{\char13}s matrix. In particular, we have shown that $g$ drops below unity by increasing the scattering strength in the system and that the level spacing switches from level repulsion, well-described by the Ginibre distribution, to level clustering with a Poisson distribution. Consistently, by estimating the localization length $\xi$ we found that $\xi/L$ is very large at low optical density, while it becomes smaller than one at the larger $\rho\lambda^2$ values that characterize the discovered localization transition. Finally, by studying the behavior of the DOS we have shown the formation of spectral gaps at large optical density and of spatially localized scattering resonances that appear where the DOS is relatively small. This behavior suggests that the localization phenomenon in the pinwheel array is driven by the suppression of the DOS, similarly to random systems \cite{John,John2}. Our findings reveal the importance of hyperuniform deterministic aperiodic structures with isotropic $k$-space for the engineering of wave localization phenomena that can be utilized to achieve enhanced light-matter interaction and novel active nanophotonic platforms.

%%%%%%%%%%%%%%%%%%%%%%%%%%%%%%%%%%%%%%%%%%%%%%%%%%%%%%%%%%
%%%%%%%%%%%%%%%%%%		  Acknowledgments              		   %%%%%%%%%%%%%%%%%%%%
%%%%%%%%%%%%%%%%%%%%%%%%%%%%%%%%%%%%%%%%%%%%%%%%%%%%%%%%%%
\begin{acknowledgments}
L.D.N. acknowledges the support from the Army Research Laboratory under Cooperative Agreement Number W911NF-12-2-0023.
\end{acknowledgments}
%\section*{Author contributions}
%F.S. performed numerical calculations, data analysis, and organized the results with the help of Y. C. S.G. performed the experiments together with F.S. W.A.B. fabricated the samples. L.D.N. conceived and \rev{led} the work. F.S. and L.D.N. wrote the manuscript.
%%%%%%%%%%%%%%%%%%%%%%%%%%%%%%%%%%%%%%%%%%%%%%%%%%%%%%%%%%
%%%%%%%%%%%%%%%%%%		        Bibilo              		   %%%%%%%%%%%%%%%%%%%%
% ==================================  Fig1  ==========================================================
%%%%%%%%%%%%%%%%%%%%%%%%%%%%%%%%%%%%%%%%%%%%%%%%%%%%%%%%%%%%%%%%%%%%%%%%%%%%%%%%%%%%%%%%%%%%%%%%%%%%%%%%%%%%%%%%%%%%%%%%%%%%%%%%%%%%%%%%%%%%%%%%%%%%%%%%%%
%%%%%%%%%%%%%%%%%%%%%%%%%%%%%%%%%%%%%%%%%%%%%%%%%%%%%%%%%%%%%%%%%%%%%%%%%%%%%%%%%%%%%%%%%%%%%%%%%%%%%%%%%%%%%%%%%%%%%%%%%%%%%%%%%%%%%%%%%%%%%%%%%%%%%%%%%%
% ==================================  Fig2  ==========================================================
%%%%%%%%%%%%%%%%%%%%%%%%%%%%%%%%%%%%%%%%%%%%%%%%%%%%%%%%%%%%%%%%%%%%%%%%%%%%%%%%%%%%%%%%%%%%%%%%%%%%%%%%%%%%%%%%%%%%%%%%%%%%%%%%%%%%%%%%%%%%%%%%%%%%%%%%%%
%%%%%%%%%%%%%%%%%%%%%%%%%%%%%%%%%%%%%%%%%%%%%%%%%%%%%%%%%%%%%%%%%%%%%%%%%%%%%%%%%%%%%%%%%%%%%%%%%%%%%%%%%%%%%%%%%%%%%%%%%%%%%%%%%%%%%%%%%%%%%%%%%%%%%%%%%%
% ==================================  Fig3  ==========================================================
%%%%%%%%%%%%%%%%%%%%%%%%%%%%%%%%%%%%%%%%%%%%%%%%%%%%%%%%%%%%%%%%%%%%%%%%%%%%%%%%%%%%%%%%%%%%%%%%%%%%%%%%%%%%%%%%%%%%%%%%%%%%%%%%%%%%%%%%%%%%%%%%%%%%%%%%%%
%%%%%%%%%%%%%%%%%%%%%%%%%%%%%%%%%%%%%%%%%%%%%%%%%%%%%%%%%%%%%%%%%%%%%%%%%%%%%%%%%%%%%%%%%%%%%%%%%%%%%%%%%%%%%%%%%%%%%%%%%%%%%%%%%%%%%%%%%%%%%%%%%%%%%%%%%%
% ==================================  Fig4  ==========================================================
%%%%%%%%%%%%%%%%%%%%%%%%%%%%%%%%%%%%%%%%%%%%%%%%%%%%%%%%%%%%%%%%%%%%%%%%%%%%%%%%%%%%%%%%%%%%%%%%%%%%%%%%%%%%%%%%%%%%%%%%%%%%%%%%%%%%%%%%%%%%%%%%%%%%%%%%%%
%%%%%%%%%%%%%%%%%%%%%%%%%%%%%%%%%%%%%%%%%%%%%%%%%%%%%%%%%%%%%%%%%%%%%%%%%%%%%%%%%%%%%%%%%%%%%%%%%%%%%%%%%%%%%%%%%%%%%%%%%%%%%%%%%%%%%%%%%%%%%%%%%%%%%%%%%%


\begin{thebibliography}{67}%
\makeatletter
\providecommand \@ifxundefined [1]{%
 \@ifx{#1\undefined}
}%
\providecommand \@ifnum [1]{%
 \ifnum #1\expandafter \@firstoftwo
 \else \expandafter \@secondoftwo
 \fi
}%
\providecommand \@ifx [1]{%
 \ifx #1\expandafter \@firstoftwo
 \else \expandafter \@secondoftwo
 \fi
}%
\providecommand \natexlab [1]{#1}%
\providecommand \emph  [1]{``#1''}%
\providecommand \bibnamefont  [1]{#1}%
\providecommand \bibfnamefont [1]{#1}%
\providecommand \citenamefont [1]{#1}%
\providecommand \href@noop [0]{\@secondoftwo}%
\providecommand \href [0]{\begingroup \@sanitize@url \@href}%
\providecommand \@href[1]{\@@startlink{#1}\@@href}%
\providecommand \@@href[1]{\endgroup#1\@@endlink}%
\providecommand \@sanitize@url [0]{\catcode `\\12\catcode `\$12\catcode
  `\&12\catcode `\#12\catcode `\^12\catcode `\_12\catcode `\%12\relax}%
\providecommand \@@startlink[1]{}%
\providecommand \@@endlink[0]{}%
\providecommand \url  [0]{\begingroup\@sanitize@url \@url }%
\providecommand \@url [1]{\endgroup\@href {#1}{\urlprefix }}%
\providecommand \urlprefix  [0]{URL }%
\providecommand \Eprint [0]{\href }%
\providecommand \doibase [0]{http://dx.doi.org/}%
\providecommand \selectlanguage [0]{\@gobble}%
\providecommand \bibinfo  [0]{\@secondoftwo}%
\providecommand \bibfield  [0]{\@secondoftwo}%
\providecommand \translation [1]{[#1]}%
\providecommand \BibitemOpen [0]{}%
\providecommand \bibitemStop [0]{}%
\providecommand \bibitemNoStop [0]{.\EOS\space}%
\providecommand \EOS [0]{\spacefactor3000\relax}%
\providecommand \BibitemShut  [1]{\csname bibitem#1\endcsname}%
\let\auto@bib@innerbib\@empty
%</preamble>
\bibitem [{\citenamefont {Florescu}\ \emph {et~al.}(2009)\citenamefont
  {Florescu}, \citenamefont {Torquato},\ and\ \citenamefont
  {Steinhardt}}]{Florescu}%
  \BibitemOpen
  \bibfield  {author} {\bibinfo {author} {\bibfnamefont {M.}~\bibnamefont
  {Florescu}}, \bibinfo {author} {\bibfnamefont {S.}~\bibnamefont {Torquato}},
  \ and\ \bibinfo {author} {\bibfnamefont {P.~J.}\ \bibnamefont {Steinhardt}},\
  }\bibfield  {title} {\emph {\bibinfo {title} {Designer disordered materials
  with large, complete photonic band gaps},}\ }\href@noop {} {\bibfield
  {journal} {\bibinfo  {journal} {Proceedings of the National Academy of
  Sciences}\ }\textbf {\bibinfo {volume} {106}},\ \bibinfo {pages} {20658}
  (\bibinfo {year} {2009})}\BibitemShut {NoStop}%
\bibitem [{\citenamefont {Pollard}\ and\ \citenamefont
  {Parker}(2009)}]{Pollard}%
  \BibitemOpen
  \bibfield  {author} {\bibinfo {author} {\bibfnamefont {M.~E.}\ \bibnamefont
  {Pollard}}\ and\ \bibinfo {author} {\bibfnamefont {G.~J.}\ \bibnamefont
  {Parker}},\ }\bibfield  {title} {\emph {\bibinfo {title} {Low-contrast
  bandgaps of a planar parabolic spiral lattice},}\ }\href@noop {} {\bibfield
  {journal} {\bibinfo  {journal} {Opt. Lett.}\ }\textbf {\bibinfo {volume}
  {34}},\ \bibinfo {pages} {2805} (\bibinfo {year} {2009})}\BibitemShut
  {NoStop}%
\bibitem [{\citenamefont {Froufe-P{\'e}rez}\ \emph {et~al.}(2016)\citenamefont
  {Froufe-P{\'e}rez}, \citenamefont {Engel}, \citenamefont {Damasceno},
  \citenamefont {Muller}, \citenamefont {Haberko}, \citenamefont {Glotzer},\
  and\ \citenamefont {Scheffold}}]{Froufe}%
  \BibitemOpen
  \bibfield  {author} {\bibinfo {author} {\bibfnamefont {L.~S.}\ \bibnamefont
  {Froufe-P{\'e}rez}}, \bibinfo {author} {\bibfnamefont {M.}~\bibnamefont
  {Engel}}, \bibinfo {author} {\bibfnamefont {P.~F.}\ \bibnamefont
  {Damasceno}}, \bibinfo {author} {\bibfnamefont {N.}~\bibnamefont {Muller}},
  \bibinfo {author} {\bibfnamefont {J.}~\bibnamefont {Haberko}}, \bibinfo
  {author} {\bibfnamefont {S.~C.}\ \bibnamefont {Glotzer}}, \ and\ \bibinfo
  {author} {\bibfnamefont {F.}~\bibnamefont {Scheffold}},\ }\bibfield  {title}
  {\emph {\bibinfo {title} {Role of short-range order and hyperuniformity in
  the formation of band gaps in disordered photonic materials},}\ }\href@noop
  {} {\bibfield  {journal} {\bibinfo  {journal} {Phys. Rev. Lett.}\ }\textbf
  {\bibinfo {volume} {117}},\ \bibinfo {pages} {053902} (\bibinfo {year}
  {2016})}\BibitemShut {NoStop}%
\bibitem [{\citenamefont {Wiesmann}\ \emph {et~al.}(2009)\citenamefont
  {Wiesmann}, \citenamefont {Bergenek}, \citenamefont {Linder},\ and\
  \citenamefont {Schwarz}}]{Wiesmann}%
  \BibitemOpen
  \bibfield  {author} {\bibinfo {author} {\bibfnamefont {C.}~\bibnamefont
  {Wiesmann}}, \bibinfo {author} {\bibfnamefont {K.}~\bibnamefont {Bergenek}},
  \bibinfo {author} {\bibfnamefont {N.}~\bibnamefont {Linder}}, \ and\ \bibinfo
  {author} {\bibfnamefont {U.~T.}\ \bibnamefont {Schwarz}},\ }\bibfield
  {title} {\emph {\bibinfo {title} {Photonic crystal LEDs--designing light
  extraction},}\ }\href@noop {} {\bibfield  {journal} {\bibinfo  {journal}
  {Laser \& Photonics Reviews}\ }\textbf {\bibinfo {volume} {3}},\ \bibinfo
  {pages} {262} (\bibinfo {year} {2009})}\BibitemShut {NoStop}%
\bibitem [{\citenamefont {Wierer}\ \emph {et~al.}(2009)\citenamefont {Wierer},
  \citenamefont {David},\ and\ \citenamefont {Megens}}]{Wierer}%
  \BibitemOpen
  \bibfield  {author} {\bibinfo {author} {\bibfnamefont {J.~J.}\ \bibnamefont
  {Wierer}}, \bibinfo {author} {\bibfnamefont {A.}~\bibnamefont {David}}, \
  and\ \bibinfo {author} {\bibfnamefont {M.~M.}\ \bibnamefont {Megens}},\
  }\bibfield  {title} {\emph {\bibinfo {title} {III-nitride photonic-crystal
  light-emitting diodes with high extraction efficiency},}\ }\href@noop {}
  {\bibfield  {journal} {\bibinfo  {journal} {Nature Photonics}\ }\textbf
  {\bibinfo {volume} {3}},\ \bibinfo {pages} {163} (\bibinfo {year}
  {2009})}\BibitemShut {NoStop}%
\bibitem [{\citenamefont {Lawrence}\ \emph
  {et~al.}(2012{\natexlab{a}})\citenamefont {Lawrence}, \citenamefont
  {Trevino},\ and\ \citenamefont {Dal~Negro}}]{LawrenceJAP}%
  \BibitemOpen
  \bibfield  {author} {\bibinfo {author} {\bibfnamefont {N.}~\bibnamefont
  {Lawrence}}, \bibinfo {author} {\bibfnamefont {J.}~\bibnamefont {Trevino}}, \
  and\ \bibinfo {author} {\bibfnamefont {L.}~\bibnamefont {Dal~Negro}},\
  }\bibfield  {title} {\emph {\bibinfo {title} {Aperiodic arrays of active
  nanopillars for radiation engineering},}\ }\href@noop {} {\bibfield
  {journal} {\bibinfo  {journal} {Journal of Applied Physics}\ }\textbf
  {\bibinfo {volume} {111}},\ \bibinfo {pages} {113101} (\bibinfo {year}
  {2012}{\natexlab{a}})}\BibitemShut {NoStop}%
\bibitem [{\citenamefont {Gorsky}\ \emph {et~al.}(2018)\citenamefont {Gorsky},
  \citenamefont {Zhang}, \citenamefont {Gok}, \citenamefont {Wang},
  \citenamefont {Kebede}, \citenamefont {Lenef}, \citenamefont {Raukas},\ and\
  \citenamefont {Dal~Negro}}]{Gorsky1}%
  \BibitemOpen
  \bibfield  {author} {\bibinfo {author} {\bibfnamefont {S.}~\bibnamefont
  {Gorsky}}, \bibinfo {author} {\bibfnamefont {R.}~\bibnamefont {Zhang}},
  \bibinfo {author} {\bibfnamefont {A.}~\bibnamefont {Gok}}, \bibinfo {author}
  {\bibfnamefont {R.}~\bibnamefont {Wang}}, \bibinfo {author} {\bibfnamefont
  {K.}~\bibnamefont {Kebede}}, \bibinfo {author} {\bibfnamefont
  {A.}~\bibnamefont {Lenef}}, \bibinfo {author} {\bibfnamefont
  {M.}~\bibnamefont {Raukas}}, \ and\ \bibinfo {author} {\bibfnamefont
  {L.}~\bibnamefont {Dal~Negro}},\ }\bibfield  {title} {\emph {\bibinfo {title}
  {Directional light emission enhancement from LED-phosphor converters using
  dielectric Vogel spiral arrays},}\ }\href@noop {} {\bibfield  {journal}
  {\bibinfo  {journal} {APL Photonics}\ }\textbf {\bibinfo {volume} {3}},\
  \bibinfo {pages} {126103} (\bibinfo {year} {2018})}\BibitemShut {NoStop}%
\bibitem [{\citenamefont {Gorsky}\ \emph {et~al.}(2019)\citenamefont {Gorsky},
  \citenamefont {Britton}, \citenamefont {Chen}, \citenamefont {Montaner},
  \citenamefont {Lenef}, \citenamefont {Raukas},\ and\ \citenamefont
  {Dal~Negro}}]{Gorsky}%
  \BibitemOpen
  \bibfield  {author} {\bibinfo {author} {\bibfnamefont {S.}~\bibnamefont
  {Gorsky}}, \bibinfo {author} {\bibfnamefont {W.}~\bibnamefont {Britton}},
  \bibinfo {author} {\bibfnamefont {Y.}~\bibnamefont {Chen}}, \bibinfo {author}
  {\bibfnamefont {J.}~\bibnamefont {Montaner}}, \bibinfo {author}
  {\bibfnamefont {A.}~\bibnamefont {Lenef}}, \bibinfo {author} {\bibfnamefont
  {M.}~\bibnamefont {Raukas}}, \ and\ \bibinfo {author} {\bibfnamefont
  {L.}~\bibnamefont {Dal~Negro}},\ }\bibfield  {title} {\emph {\bibinfo {title}
  {Engineered hyperuniformity for directional light extraction},}\ }\href@noop
  {} {\bibfield  {journal} {\bibinfo  {journal} {APL Photonics}\ }\textbf
  {\bibinfo {volume} {4}},\ \bibinfo {pages} {110801} (\bibinfo {year}
  {2019})}\BibitemShut {NoStop}%
\bibitem [{\citenamefont {Notomi}\ \emph {et~al.}(2004)\citenamefont {Notomi},
  \citenamefont {Suzuki}, \citenamefont {Tamamura},\ and\ \citenamefont
  {Edagawa}}]{Notomi}%
  \BibitemOpen
  \bibfield  {author} {\bibinfo {author} {\bibfnamefont {M.}~\bibnamefont
  {Notomi}}, \bibinfo {author} {\bibfnamefont {H.}~\bibnamefont {Suzuki}},
  \bibinfo {author} {\bibfnamefont {T.}~\bibnamefont {Tamamura}}, \ and\
  \bibinfo {author} {\bibfnamefont {K.}~\bibnamefont {Edagawa}},\ }\bibfield
  {title} {\emph {\bibinfo {title} {Lasing action due to the two-dimensional
  quasiperiodicity of photonic quasicrystals with a Penrose lattice},}\
  }\href@noop {} {\bibfield  {journal} {\bibinfo  {journal} {Phys. Rev. Lett.}\
  }\textbf {\bibinfo {volume} {92}},\ \bibinfo {pages} {123906} (\bibinfo
  {year} {2004})}\BibitemShut {NoStop}%
\bibitem [{\citenamefont {Lee}\ \emph {et~al.}(2011)\citenamefont {Lee},
  \citenamefont {Forestiere}, \citenamefont {Pasquale}, \citenamefont
  {Trevino}, \citenamefont {Walsh}, \citenamefont {Galli}, \citenamefont
  {Romagnoli},\ and\ \citenamefont {Dal~Negro}}]{Lee}%
  \BibitemOpen
  \bibfield  {author} {\bibinfo {author} {\bibfnamefont {S.~Y.}\ \bibnamefont
  {Lee}}, \bibinfo {author} {\bibfnamefont {C.}~\bibnamefont {Forestiere}},
  \bibinfo {author} {\bibfnamefont {A.~J.}\ \bibnamefont {Pasquale}}, \bibinfo
  {author} {\bibfnamefont {J.}~\bibnamefont {Trevino}}, \bibinfo {author}
  {\bibfnamefont {G.}~\bibnamefont {Walsh}}, \bibinfo {author} {\bibfnamefont
  {P.}~\bibnamefont {Galli}}, \bibinfo {author} {\bibfnamefont
  {M.}~\bibnamefont {Romagnoli}}, \ and\ \bibinfo {author} {\bibfnamefont
  {L.}~\bibnamefont {Dal~Negro}},\ }\bibfield  {title} {\emph {\bibinfo {title}
  {Plasmon-enhanced structural coloration of metal films with isotropic
  Pinwheel nanoparticle arrays},}\ }\href@noop {} {\bibfield  {journal}
  {\bibinfo  {journal} {Opt. Exp.}\ }\textbf {\bibinfo {volume} {19}},\
  \bibinfo {pages} {23818} (\bibinfo {year} {2011})}\BibitemShut {NoStop}%
\bibitem [{\citenamefont {Boriskina}\ \emph {et~al.}(2010)\citenamefont
  {Boriskina}, \citenamefont {Lee}, \citenamefont {Amsden}, \citenamefont
  {Omenetto},\ and\ \citenamefont {Dal~Negro}}]{Boriskina}%
  \BibitemOpen
  \bibfield  {author} {\bibinfo {author} {\bibfnamefont {S.~V.}\ \bibnamefont
  {Boriskina}}, \bibinfo {author} {\bibfnamefont {S.~Y.}\ \bibnamefont {Lee}},
  \bibinfo {author} {\bibfnamefont {J.~J.}\ \bibnamefont {Amsden}}, \bibinfo
  {author} {\bibfnamefont {F.~G.}\ \bibnamefont {Omenetto}}, \ and\ \bibinfo
  {author} {\bibfnamefont {L.}~\bibnamefont {Dal~Negro}},\ }\bibfield  {title}
  {\emph {\bibinfo {title} {Formation of colorimetric fingerprints on
  nano-patterned deterministic aperiodic surfaces},}\ }\href@noop {} {\bibfield
   {journal} {\bibinfo  {journal} {Opt. Exp.}\ }\textbf {\bibinfo {volume}
  {18}},\ \bibinfo {pages} {14568} (\bibinfo {year} {2010})}\BibitemShut
  {NoStop}%
\bibitem [{\citenamefont {Trojak}\ \emph {et~al.}(2021)\citenamefont {Trojak},
  \citenamefont {Gorsky}, \citenamefont {Murray}, \citenamefont {Sgrignuoli},
  \citenamefont {Pinheiro}, \citenamefont {Dal~Negro},\ and\ \citenamefont
  {Sapienza}}]{Trojak}%
  \BibitemOpen
  \bibfield  {author} {\bibinfo {author} {\bibfnamefont {O.~J.}\ \bibnamefont
  {Trojak}}, \bibinfo {author} {\bibfnamefont {S.}~\bibnamefont {Gorsky}},
  \bibinfo {author} {\bibfnamefont {C.}~\bibnamefont {Murray}}, \bibinfo
  {author} {\bibfnamefont {F.}~\bibnamefont {Sgrignuoli}}, \bibinfo {author}
  {\bibfnamefont {F.~A.}\ \bibnamefont {Pinheiro}}, \bibinfo {author}
  {\bibfnamefont {L.}~\bibnamefont {Dal~Negro}}, \ and\ \bibinfo {author}
  {\bibfnamefont {L.}~\bibnamefont {Sapienza}},\ }\bibfield  {title} {\emph
  {\bibinfo {title} {Cavity-enhanced light--matter interaction in Vogel-spiral
  devices as a platform for quantum photonics},}\ }\href@noop {} {\bibfield
  {journal} {\bibinfo  {journal} {Appl. Phys. Lett.}\ }\textbf {\bibinfo
  {volume} {118}},\ \bibinfo {pages} {011103} (\bibinfo {year}
  {2021})}\BibitemShut {NoStop}%
\bibitem [{\citenamefont {Trojak}\ \emph {et~al.}(2020)\citenamefont {Trojak},
  \citenamefont {Gorsky}, \citenamefont {Sgrignuoli}, \citenamefont {Pinheiro},
  \citenamefont {Park}, \citenamefont {Song}, \citenamefont {Dal~Negro},\ and\
  \citenamefont {Sapienza}}]{Trojak2}%
  \BibitemOpen
  \bibfield  {author} {\bibinfo {author} {\bibfnamefont {O.~J.}\ \bibnamefont
  {Trojak}}, \bibinfo {author} {\bibfnamefont {S.}~\bibnamefont {Gorsky}},
  \bibinfo {author} {\bibfnamefont {F.}~\bibnamefont {Sgrignuoli}}, \bibinfo
  {author} {\bibfnamefont {F.~A.}\ \bibnamefont {Pinheiro}}, \bibinfo {author}
  {\bibfnamefont {S.-I.}\ \bibnamefont {Park}}, \bibinfo {author}
  {\bibfnamefont {J.~D.}\ \bibnamefont {Song}}, \bibinfo {author}
  {\bibfnamefont {L.}~\bibnamefont {Dal~Negro}}, \ and\ \bibinfo {author}
  {\bibfnamefont {L.}~\bibnamefont {Sapienza}},\ }\bibfield  {title} {\emph
  {\bibinfo {title} {Cavity quantum electro-dynamics with solid-state emitters
  in aperiodic nano-photonic spiral devices},}\ }\href@noop {} {\bibfield
  {journal} {\bibinfo  {journal} {Appl. Phys. Lett.}\ }\textbf {\bibinfo
  {volume} {117}},\ \bibinfo {pages} {124006} (\bibinfo {year}
  {2020})}\BibitemShut {NoStop}%
\bibitem [{\citenamefont {Torquato}(2018)}]{TorquatoReview}%
  \BibitemOpen
  \bibfield  {author} {\bibinfo {author} {\bibfnamefont {S.}~\bibnamefont
  {Torquato}},\ }\bibfield  {title} {\emph {\bibinfo {title} {Hyperuniform
  states of matter},}\ }\href@noop {} {\bibfield  {journal} {\bibinfo
  {journal} {Phys. Rep.}\ }\textbf {\bibinfo {volume} {745}},\ \bibinfo {pages}
  {1} (\bibinfo {year} {2018})}\BibitemShut {NoStop}%
\bibitem [{\citenamefont {David}\ \emph {et~al.}(2001)\citenamefont {David},
  \citenamefont {Chelnikov},\ and\ \citenamefont {Lourtioz}}]{David}%
  \BibitemOpen
  \bibfield  {author} {\bibinfo {author} {\bibfnamefont {S.}~\bibnamefont
  {David}}, \bibinfo {author} {\bibfnamefont {A.}~\bibnamefont {Chelnikov}}, \
  and\ \bibinfo {author} {\bibfnamefont {J.-M.}\ \bibnamefont {Lourtioz}},\
  }\bibfield  {title} {\emph {\bibinfo {title} {Isotropic photonic structures:
  Archimedean-like tilings and quasi-crystals},}\ }\href@noop {} {\bibfield
  {journal} {\bibinfo  {journal} {IEEE Journal of Quantum Electronics}\
  }\textbf {\bibinfo {volume} {37}},\ \bibinfo {pages} {1427} (\bibinfo {year}
  {2001})}\BibitemShut {NoStop}%
\bibitem [{\citenamefont {Hagelstein}\ and\ \citenamefont
  {Denison}(1999)}]{Hagelstein}%
  \BibitemOpen
  \bibfield  {author} {\bibinfo {author} {\bibfnamefont {P.~L.}\ \bibnamefont
  {Hagelstein}}\ and\ \bibinfo {author} {\bibfnamefont {D.~R.}\ \bibnamefont
  {Denison}},\ }\bibfield  {title} {\emph {\bibinfo {title} {Nearly isotropic
  photonic bandgap structures in two dimensions},}\ }\href@noop {} {\bibfield
  {journal} {\bibinfo  {journal} {Opt. Lett.}\ }\textbf {\bibinfo {volume}
  {24}},\ \bibinfo {pages} {708} (\bibinfo {year} {1999})}\BibitemShut
  {NoStop}%
\bibitem [{\citenamefont {Guo}\ \emph {et~al.}(2017)\citenamefont {Guo},
  \citenamefont {Du}, \citenamefont {Osorio},\ and\ \citenamefont
  {Koenderink}}]{Guo}%
  \BibitemOpen
  \bibfield  {author} {\bibinfo {author} {\bibfnamefont {K.}~\bibnamefont
  {Guo}}, \bibinfo {author} {\bibfnamefont {M.}~\bibnamefont {Du}}, \bibinfo
  {author} {\bibfnamefont {C.~I.}\ \bibnamefont {Osorio}}, \ and\ \bibinfo
  {author} {\bibfnamefont {A.~F.}\ \bibnamefont {Koenderink}},\ }\bibfield
  {title} {\emph {\bibinfo {title} {Broadband light scattering and
  photoluminescence enhancement from plasmonic Vogel's golden spirals},}\
  }\href@noop {} {\bibfield  {journal} {\bibinfo  {journal} {Laser \& Photonics
  Reviews}\ }\textbf {\bibinfo {volume} {11}},\ \bibinfo {pages} {1600235}
  (\bibinfo {year} {2017})}\BibitemShut {NoStop}%
\bibitem [{\citenamefont {Trevino}\ \emph {et~al.}(2011)\citenamefont
  {Trevino}, \citenamefont {Cao},\ and\ \citenamefont {Dal~Negro}}]{Trevino}%
  \BibitemOpen
  \bibfield  {author} {\bibinfo {author} {\bibfnamefont {J.}~\bibnamefont
  {Trevino}}, \bibinfo {author} {\bibfnamefont {H.}~\bibnamefont {Cao}}, \ and\
  \bibinfo {author} {\bibfnamefont {L.}~\bibnamefont {Dal~Negro}},\ }\bibfield
  {title} {\emph {\bibinfo {title} {Circularly symmetric light scattering from
  nanoplasmonic spirals},}\ }\href@noop {} {\bibfield  {journal} {\bibinfo
  {journal} {Nano Lett.}\ }\textbf {\bibinfo {volume} {11}},\ \bibinfo {pages}
  {2008} (\bibinfo {year} {2011})}\BibitemShut {NoStop}%
\bibitem [{\citenamefont {Dal~Negro}\ \emph {et~al.}(2012)\citenamefont
  {Dal~Negro}, \citenamefont {Lawrence},\ and\ \citenamefont
  {Trevino}}]{DalNegroVogel}%
  \BibitemOpen
  \bibfield  {author} {\bibinfo {author} {\bibfnamefont {L.}~\bibnamefont
  {Dal~Negro}}, \bibinfo {author} {\bibfnamefont {N.}~\bibnamefont {Lawrence}},
  \ and\ \bibinfo {author} {\bibfnamefont {J.}~\bibnamefont {Trevino}},\
  }\bibfield  {title} {\emph {\bibinfo {title} {Analytical light scattering and
  orbital angular momentum spectra of arbitrary Vogel spirals},}\ }\href@noop
  {} {\bibfield  {journal} {\bibinfo  {journal} {Opt. Exp.}\ }\textbf {\bibinfo
  {volume} {20}},\ \bibinfo {pages} {18209} (\bibinfo {year}
  {2012})}\BibitemShut {NoStop}%
\bibitem [{\citenamefont {Lawrence}\ \emph
  {et~al.}(2012{\natexlab{b}})\citenamefont {Lawrence}, \citenamefont
  {Trevino},\ and\ \citenamefont {Dal~Negro}}]{Lawrence}%
  \BibitemOpen
  \bibfield  {author} {\bibinfo {author} {\bibfnamefont {N.}~\bibnamefont
  {Lawrence}}, \bibinfo {author} {\bibfnamefont {J.}~\bibnamefont {Trevino}}, \
  and\ \bibinfo {author} {\bibfnamefont {L.}~\bibnamefont {Dal~Negro}},\
  }\bibfield  {title} {\emph {\bibinfo {title} {Control of optical orbital
  angular momentum by Vogel spiral arrays of metallic nanoparticles},}\
  }\href@noop {} {\bibfield  {journal} {\bibinfo  {journal} {Opt. Lett.}\
  }\textbf {\bibinfo {volume} {37}},\ \bibinfo {pages} {5076} (\bibinfo {year}
  {2012}{\natexlab{b}})}\BibitemShut {NoStop}%
\bibitem [{\citenamefont {Pierro}\ \emph {et~al.}(2005)\citenamefont {Pierro},
  \citenamefont {Galdi}, \citenamefont {Castaldi}, \citenamefont {Pinto},\ and\
  \citenamefont {Felsen}}]{Pierro}%
  \BibitemOpen
  \bibfield  {author} {\bibinfo {author} {\bibfnamefont {V.}~\bibnamefont
  {Pierro}}, \bibinfo {author} {\bibfnamefont {V.}~\bibnamefont {Galdi}},
  \bibinfo {author} {\bibfnamefont {G.}~\bibnamefont {Castaldi}}, \bibinfo
  {author} {\bibfnamefont {I.~M.}\ \bibnamefont {Pinto}}, \ and\ \bibinfo
  {author} {\bibfnamefont {L.~B.}\ \bibnamefont {Felsen}},\ }\bibfield  {title}
  {\emph {\bibinfo {title} {Radiation properties of planar antenna arrays based
  on certain categories of aperiodic tilings},}\ }\href@noop {} {\bibfield
  {journal} {\bibinfo  {journal} {IEEE transactions on antennas and
  propagation}\ }\textbf {\bibinfo {volume} {53}},\ \bibinfo {pages} {635}
  (\bibinfo {year} {2005})}\BibitemShut {NoStop}%
\bibitem [{\citenamefont {Radin}(1994)}]{Radin1994}%
  \BibitemOpen
  \bibfield  {author} {\bibinfo {author} {\bibfnamefont {C.}~\bibnamefont
  {Radin}},\ }\bibfield  {title} {\emph {\bibinfo {title} {The pinwheel tilings
  of the plane},}\ }\href@noop {} {\bibfield  {journal} {\bibinfo  {journal}
  {Annals of Mathematics}\ }\textbf {\bibinfo {volume} {139}},\ \bibinfo
  {pages} {661} (\bibinfo {year} {1994})}\BibitemShut {NoStop}%
\bibitem [{\citenamefont {Radin}(1999)}]{RadinBook}%
  \BibitemOpen
  \bibfield  {author} {\bibinfo {author} {\bibfnamefont {C.}~\bibnamefont
  {Radin}},\ }\href@noop {} {\emph {\bibinfo {title} {Miles of tiles}}},\
  Vol.~\bibinfo {volume} {1}\ (\bibinfo  {publisher} {American Mathematical
  Soc.},\ \bibinfo {year} {1999})\BibitemShut {NoStop}%
\bibitem [{\citenamefont {Aubry}\ \emph {et~al.}(2020)\citenamefont {Aubry},
  \citenamefont {Froufe-P{\'e}rez}, \citenamefont {Kuhl}, \citenamefont
  {Legrand}, \citenamefont {Scheffold},\ and\ \citenamefont
  {Mortessagne}}]{Aubry}%
  \BibitemOpen
  \bibfield  {author} {\bibinfo {author} {\bibfnamefont {G.~J.}\ \bibnamefont
  {Aubry}}, \bibinfo {author} {\bibfnamefont {L.~S.}\ \bibnamefont
  {Froufe-P{\'e}rez}}, \bibinfo {author} {\bibfnamefont {U.}~\bibnamefont
  {Kuhl}}, \bibinfo {author} {\bibfnamefont {O.}~\bibnamefont {Legrand}},
  \bibinfo {author} {\bibfnamefont {F.}~\bibnamefont {Scheffold}}, \ and\
  \bibinfo {author} {\bibfnamefont {F.}~\bibnamefont {Mortessagne}},\
  }\bibfield  {title} {\emph {\bibinfo {title} {Experimental tuning of
  transport regimes in hyperuniform disordered photonic materials},}\
  }\href@noop {} {\bibfield  {journal} {\bibinfo  {journal} {Phys. Rev. Lett.}\
  }\textbf {\bibinfo {volume} {125}},\ \bibinfo {pages} {127402} (\bibinfo
  {year} {2020})}\BibitemShut {NoStop}%
\bibitem [{\citenamefont {Froufe-P{\'e}rez}\ \emph {et~al.}(2017)\citenamefont
  {Froufe-P{\'e}rez}, \citenamefont {Engel}, \citenamefont {S{\'a}enz},\ and\
  \citenamefont {Scheffold}}]{FroufePNAS}%
  \BibitemOpen
  \bibfield  {author} {\bibinfo {author} {\bibfnamefont {L.~S.}\ \bibnamefont
  {Froufe-P{\'e}rez}}, \bibinfo {author} {\bibfnamefont {M.}~\bibnamefont
  {Engel}}, \bibinfo {author} {\bibfnamefont {J.~J.}\ \bibnamefont
  {S{\'a}enz}}, \ and\ \bibinfo {author} {\bibfnamefont {F.}~\bibnamefont
  {Scheffold}},\ }\bibfield  {title} {\emph {\bibinfo {title} {Band gap
  formation and Anderson localization in disordered photonic materials with
  structural correlations},}\ }\href@noop {} {\bibfield  {journal} {\bibinfo
  {journal} {Proceedings of the National Academy of Sciences}\ }\textbf
  {\bibinfo {volume} {114}},\ \bibinfo {pages} {9570} (\bibinfo {year}
  {2017})}\BibitemShut {NoStop}%
\bibitem [{\citenamefont {Lagendijk}\ and\ \citenamefont
  {Van~Tiggelen}(1996)}]{Lagendijk}%
  \BibitemOpen
  \bibfield  {author} {\bibinfo {author} {\bibfnamefont {A.}~\bibnamefont
  {Lagendijk}}\ and\ \bibinfo {author} {\bibfnamefont {B.~A.}\ \bibnamefont
  {Van~Tiggelen}},\ }\bibfield  {title} {\emph {\bibinfo {title} {Resonant
  multiple scattering of light},}\ }\href@noop {} {\bibfield  {journal}
  {\bibinfo  {journal} {Phys. Rep}\ }\textbf {\bibinfo {volume} {270}},\
  \bibinfo {pages} {143} (\bibinfo {year} {1996})}\BibitemShut {NoStop}%
\bibitem [{\citenamefont {Rusek}\ and\ \citenamefont
  {Or{\l}owski}(1995)}]{RusekPRE2D}%
  \BibitemOpen
  \bibfield  {author} {\bibinfo {author} {\bibfnamefont {M.}~\bibnamefont
  {Rusek}}\ and\ \bibinfo {author} {\bibfnamefont {A.}~\bibnamefont
  {Or{\l}owski}},\ }\bibfield  {title} {\emph {\bibinfo {title} {Analytical
  approach to localization of electromagnetic waves in two-dimensional random
  media},}\ }\href@noop {} {\bibfield  {journal} {\bibinfo  {journal} {Phys.
  Rev. E}\ }\textbf {\bibinfo {volume} {51}},\ \bibinfo {pages} {R2763}
  (\bibinfo {year} {1995})}\BibitemShut {NoStop}%
\bibitem [{\citenamefont {John}(1987)}]{John}%
  \BibitemOpen
  \bibfield  {author} {\bibinfo {author} {\bibfnamefont {S.}~\bibnamefont
  {John}},\ }\bibfield  {title} {\emph {\bibinfo {title} {Strong localization
  of photons in certain disordered dielectric superlattices},}\ }\href@noop {}
  {\bibfield  {journal} {\bibinfo  {journal} {Phys. Rev. Lett.}\ }\textbf
  {\bibinfo {volume} {58}},\ \bibinfo {pages} {2486} (\bibinfo {year}
  {1987})}\BibitemShut {NoStop}%
\bibitem [{\citenamefont {John}(1991)}]{John2}%
  \BibitemOpen
  \bibfield  {author} {\bibinfo {author} {\bibfnamefont {S.}~\bibnamefont
  {John}},\ }\bibfield  {title} {\emph {\bibinfo {title} {Localization of
  light},}\ }\href@noop {} {\bibfield  {journal} {\bibinfo  {journal} {Phys.
  Today}\ }\textbf {\bibinfo {volume} {44}},\ \bibinfo {pages} {32} (\bibinfo
  {year} {1991})}\BibitemShut {NoStop}%
\bibitem [{\citenamefont {Skipetrov}(2020)}]{Skipetrov2020}%
  \BibitemOpen
  \bibfield  {author} {\bibinfo {author} {\bibfnamefont {S.}~\bibnamefont
  {Skipetrov}},\ }\bibfield  {title} {\emph {\bibinfo {title} {Localization of
  light in a three-dimensional disordered crystal of atoms},}\ }\href@noop {}
  {\bibfield  {journal} {\bibinfo  {journal} {Phys. Rev. B}\ }\textbf {\bibinfo
  {volume} {102}},\ \bibinfo {pages} {134206} (\bibinfo {year}
  {2020})}\BibitemShut {NoStop}%
\bibitem [{\citenamefont {Baake}\ and\ \citenamefont {Grimm}(2013)}]{Baake}%
  \BibitemOpen
  \bibfield  {author} {\bibinfo {author} {\bibfnamefont {M.}~\bibnamefont
  {Baake}}\ and\ \bibinfo {author} {\bibfnamefont {U.}~\bibnamefont {Grimm}},\
  }\href@noop {} {\emph {\bibinfo {title} {Aperiodic order}}},\ Vol.~\bibinfo
  {volume} {1}\ (\bibinfo  {publisher} {Cambridge University Press, Cambridge
  (UK)},\ \bibinfo {year} {2013})\BibitemShut {NoStop}%
\bibitem [{\citenamefont {Senechal}(1996)}]{Senechal}%
  \BibitemOpen
  \bibfield  {author} {\bibinfo {author} {\bibfnamefont {M.}~\bibnamefont
  {Senechal}},\ }\href@noop {} {\emph {\bibinfo {title} {Quasicrystals and
  geometry}}}\ (\bibinfo  {publisher} {CUP Archive},\ \bibinfo {year}
  {1996})\BibitemShut {NoStop}%
\bibitem [{\citenamefont {Moody}\ \emph {et~al.}(2006)\citenamefont {Moody},
  \citenamefont {Postnikoff},\ and\ \citenamefont {Strungaru}}]{Moody}%
  \BibitemOpen
  \bibfield  {author} {\bibinfo {author} {\bibfnamefont {R.~V.}\ \bibnamefont
  {Moody}}, \bibinfo {author} {\bibfnamefont {D.}~\bibnamefont {Postnikoff}}, \
  and\ \bibinfo {author} {\bibfnamefont {N.}~\bibnamefont {Strungaru}},\
  }\bibfield  {title} {\emph {\bibinfo {title} {Circular symmetry of pinwheel
  diffraction},}\ }in\ \href@noop {} {\emph {\bibinfo {booktitle} {Annales
  Henri Poincar{\'e}}}},\ Vol.~\bibinfo {volume} {7}\ (\bibinfo {organization}
  {Springer},\ \bibinfo {year} {2006})\ pp.\ \bibinfo {pages}
  {711--730}\BibitemShut {NoStop}%
\bibitem [{\citenamefont {Baake}\ \emph {et~al.}(2007)\citenamefont {Baake},
  \citenamefont {Frettl{\"o}h},\ and\ \citenamefont {Grimm}}]{BaakePinwh}%
  \BibitemOpen
  \bibfield  {author} {\bibinfo {author} {\bibfnamefont {M.}~\bibnamefont
  {Baake}}, \bibinfo {author} {\bibfnamefont {D.}~\bibnamefont {Frettl{\"o}h}},
  \ and\ \bibinfo {author} {\bibfnamefont {U.}~\bibnamefont {Grimm}},\
  }\bibfield  {title} {\emph {\bibinfo {title} {Pinwheel patterns and powder
  diffraction},}\ }\href@noop {} {\bibfield  {journal} {\bibinfo  {journal}
  {Philosophical Magazine}\ }\textbf {\bibinfo {volume} {87}},\ \bibinfo
  {pages} {2831} (\bibinfo {year} {2007})}\BibitemShut {NoStop}%
\bibitem [{\citenamefont {Gabrielli}\ \emph {et~al.}(2003)\citenamefont
  {Gabrielli}, \citenamefont {Jancovici}, \citenamefont {Joyce}, \citenamefont
  {Lebowitz}, \citenamefont {Pietronero},\ and\ \citenamefont
  {Labini}}]{Gabrielli}%
  \BibitemOpen
  \bibfield  {author} {\bibinfo {author} {\bibfnamefont {A.}~\bibnamefont
  {Gabrielli}}, \bibinfo {author} {\bibfnamefont {B.}~\bibnamefont
  {Jancovici}}, \bibinfo {author} {\bibfnamefont {M.}~\bibnamefont {Joyce}},
  \bibinfo {author} {\bibfnamefont {J.}~\bibnamefont {Lebowitz}}, \bibinfo
  {author} {\bibfnamefont {L.}~\bibnamefont {Pietronero}}, \ and\ \bibinfo
  {author} {\bibfnamefont {F.~S.}\ \bibnamefont {Labini}},\ }\bibfield  {title}
  {\emph {\bibinfo {title} {Generation of primordial cosmological perturbations
  from statistical mechanical models},}\ }\href@noop {} {\bibfield  {journal}
  {\bibinfo  {journal} {Physical Review D}\ }\textbf {\bibinfo {volume} {67}},\
  \bibinfo {pages} {043506} (\bibinfo {year} {2003})}\BibitemShut {NoStop}%
\bibitem [{\citenamefont {Torquato}\ and\ \citenamefont
  {Stillinger}(2003)}]{Torquato}%
  \BibitemOpen
  \bibfield  {author} {\bibinfo {author} {\bibfnamefont {S.}~\bibnamefont
  {Torquato}}\ and\ \bibinfo {author} {\bibfnamefont {F.~H.}\ \bibnamefont
  {Stillinger}},\ }\bibfield  {title} {\emph {\bibinfo {title} {Local density
  fluctuations, hyperuniformity, and order metrics},}\ }\href@noop {}
  {\bibfield  {journal} {\bibinfo  {journal} {Phys. Rev. E}\ }\textbf {\bibinfo
  {volume} {68}},\ \bibinfo {pages} {041113} (\bibinfo {year}
  {2003})}\BibitemShut {NoStop}%
\bibitem [{\citenamefont {Sgrignuoli}\ and\ \citenamefont
  {Dal~Negro}(2020)}]{Sgrignuoli_PRB2020}%
  \BibitemOpen
  \bibfield  {author} {\bibinfo {author} {\bibfnamefont {F.}~\bibnamefont
  {Sgrignuoli}}\ and\ \bibinfo {author} {\bibfnamefont {L.}~\bibnamefont
  {Dal~Negro}},\ }\bibfield  {title} {\emph {\bibinfo {title} {Subdiffusive
  light transport in three-dimensional subrandom arrays},}\ }\href@noop {}
  {\bibfield  {journal} {\bibinfo  {journal} {Phys. Rev. B}\ }\textbf {\bibinfo
  {volume} {101}},\ \bibinfo {pages} {214204} (\bibinfo {year}
  {2020})}\BibitemShut {NoStop}%
\bibitem [{\citenamefont {Haberko}\ \emph {et~al.}(2020)\citenamefont
  {Haberko}, \citenamefont {Froufe-P{\'e}rez},\ and\ \citenamefont
  {Scheffold}}]{Haberko}%
  \BibitemOpen
  \bibfield  {author} {\bibinfo {author} {\bibfnamefont {J.}~\bibnamefont
  {Haberko}}, \bibinfo {author} {\bibfnamefont {L.~S.}\ \bibnamefont
  {Froufe-P{\'e}rez}}, \ and\ \bibinfo {author} {\bibfnamefont
  {F.}~\bibnamefont {Scheffold}},\ }\bibfield  {title} {\emph {\bibinfo {title}
  {Transition from light diffusion to localization in three-dimensional
  amorphous dielectric networks near the band edge},}\ }\href@noop {}
  {\bibfield  {journal} {\bibinfo  {journal} {Nature Comm.}\ }\textbf {\bibinfo
  {volume} {11}},\ \bibinfo {pages} {1} (\bibinfo {year} {2020})}\BibitemShut
  {NoStop}%
\bibitem [{\citenamefont {Yu}\ \emph {et~al.}(2020)\citenamefont {Yu},
  \citenamefont {Qiu}, \citenamefont {Chong}, \citenamefont {Torquato},\ and\
  \citenamefont {Park}}]{Yu}%
  \BibitemOpen
  \bibfield  {author} {\bibinfo {author} {\bibfnamefont {S.}~\bibnamefont
  {Yu}}, \bibinfo {author} {\bibfnamefont {C.-W.}\ \bibnamefont {Qiu}},
  \bibinfo {author} {\bibfnamefont {Y.}~\bibnamefont {Chong}}, \bibinfo
  {author} {\bibfnamefont {S.}~\bibnamefont {Torquato}}, \ and\ \bibinfo
  {author} {\bibfnamefont {N.}~\bibnamefont {Park}},\ }\bibfield  {title}
  {\emph {\bibinfo {title} {Engineered disorder in photonics},}\ }\href@noop {}
  {\bibfield  {journal} {\bibinfo  {journal} {Nature Reviews Materials}\ ,\
  \bibinfo {pages} {1}} (\bibinfo {year} {2020})}\BibitemShut {NoStop}%
\bibitem [{\citenamefont {O{\u{g}}uz}\ \emph {et~al.}(2017)\citenamefont
  {O{\u{g}}uz}, \citenamefont {Socolar}, \citenamefont {Steinhardt},\ and\
  \citenamefont {Torquato}}]{Ouguz}%
  \BibitemOpen
  \bibfield  {author} {\bibinfo {author} {\bibfnamefont {E.~C.}\ \bibnamefont
  {O{\u{g}}uz}}, \bibinfo {author} {\bibfnamefont {J.~E.}\ \bibnamefont
  {Socolar}}, \bibinfo {author} {\bibfnamefont {P.~J.}\ \bibnamefont
  {Steinhardt}}, \ and\ \bibinfo {author} {\bibfnamefont {S.}~\bibnamefont
  {Torquato}},\ }\bibfield  {title} {\emph {\bibinfo {title} {Hyperuniformity
  of quasicrystals},}\ }\href@noop {} {\bibfield  {journal} {\bibinfo
  {journal} {Phys. Rev. B}\ }\textbf {\bibinfo {volume} {95}},\ \bibinfo
  {pages} {054119} (\bibinfo {year} {2017})}\BibitemShut {NoStop}%
\bibitem [{\citenamefont {Negro}(2021)}]{LucaBibbiaBook}%
  \BibitemOpen
  \bibfield  {author} {\bibinfo {author} {\bibfnamefont {L.~D.}\ \bibnamefont
  {Negro}},\ }\href@noop {} {\emph {\bibinfo {title} {Waves in complex
  media}}}\ (\bibinfo  {publisher} {Cambridge University Press, London (UK),
  \emph{in press}},\ \bibinfo {year} {2021})\BibitemShut {NoStop}%
\bibitem [{\citenamefont {Mehta}(2004)}]{Mehta}%
  \BibitemOpen
  \bibfield  {author} {\bibinfo {author} {\bibfnamefont {M.~L.}\ \bibnamefont
  {Mehta}},\ }\href@noop {} {\emph {\bibinfo {title} {Random matrices}}},\
  Vol.\ \bibinfo {volume} {142}\ (\bibinfo  {publisher} {Elsevier},\ \bibinfo
  {year} {2004})\BibitemShut {NoStop}%
\bibitem [{\citenamefont {Bohigas}\ \emph {et~al.}(1985)\citenamefont
  {Bohigas}, \citenamefont {Haq},\ and\ \citenamefont {Pandey}}]{Bohigas}%
  \BibitemOpen
  \bibfield  {author} {\bibinfo {author} {\bibfnamefont {O.}~\bibnamefont
  {Bohigas}}, \bibinfo {author} {\bibfnamefont {R.~U.}\ \bibnamefont {Haq}}, \
  and\ \bibinfo {author} {\bibfnamefont {A.}~\bibnamefont {Pandey}},\
  }\bibfield  {title} {\emph {\bibinfo {title} {Higher-order correlations in
  spectra of complex systems},}\ }\href@noop {} {\bibfield  {journal} {\bibinfo
   {journal} {Phys. Rev. Lett.}\ }\textbf {\bibinfo {volume} {54}},\ \bibinfo
  {pages} {1645} (\bibinfo {year} {1985})}\BibitemShut {NoStop}%
\bibitem [{\citenamefont {Torquato}\ \emph {et~al.}(2020)\citenamefont
  {Torquato}, \citenamefont {Kim},\ and\ \citenamefont
  {Klatt}}]{Torquato2020local}%
  \BibitemOpen
  \bibfield  {author} {\bibinfo {author} {\bibfnamefont {S.}~\bibnamefont
  {Torquato}}, \bibinfo {author} {\bibfnamefont {J.}~\bibnamefont {Kim}}, \
  and\ \bibinfo {author} {\bibfnamefont {M.~A.}\ \bibnamefont {Klatt}},\
  }\bibfield  {title} {\emph {\bibinfo {title} {Local Number Fluctuations in
  Hyperuniform and Nonhyperuniform Systems: Higher-Order Moments and
  Distribution Functions},}\ }\href@noop {} {\bibfield  {journal} {\bibinfo
  {journal} {arXiv preprint arXiv:2012.02358}\ } (\bibinfo {year}
  {2020})}\BibitemShut {NoStop}%
\bibitem [{\citenamefont {Leseur}\ \emph {et~al.}(2016)\citenamefont {Leseur},
  \citenamefont {Pierrat},\ and\ \citenamefont {Carminati}}]{Leseur}%
  \BibitemOpen
  \bibfield  {author} {\bibinfo {author} {\bibfnamefont {O.}~\bibnamefont
  {Leseur}}, \bibinfo {author} {\bibfnamefont {R.}~\bibnamefont {Pierrat}}, \
  and\ \bibinfo {author} {\bibfnamefont {R.}~\bibnamefont {Carminati}},\
  }\bibfield  {title} {\emph {\bibinfo {title} {High-density hyperuniform
  materials can be transparent},}\ }\href@noop {} {\bibfield  {journal}
  {\bibinfo  {journal} {Optica}\ }\textbf {\bibinfo {volume} {3}},\ \bibinfo
  {pages} {763} (\bibinfo {year} {2016})}\BibitemShut {NoStop}%
\bibitem [{\citenamefont {Dal~Negro}\ \emph {et~al.}(2019)\citenamefont
  {Dal~Negro}, \citenamefont {Chen},\ and\ \citenamefont
  {Sgrignuoli}}]{DalNegroElliptic}%
  \BibitemOpen
  \bibfield  {author} {\bibinfo {author} {\bibfnamefont {L.}~\bibnamefont
  {Dal~Negro}}, \bibinfo {author} {\bibfnamefont {Y.}~\bibnamefont {Chen}}, \
  and\ \bibinfo {author} {\bibfnamefont {F.}~\bibnamefont {Sgrignuoli}},\
  }\bibfield  {title} {\emph {\bibinfo {title} {Aperiodic Photonics of Elliptic
  Curves},}\ }\href@noop {} {\bibfield  {journal} {\bibinfo  {journal}
  {Crystals}\ }\textbf {\bibinfo {volume} {9}},\ \bibinfo {pages} {482}
  (\bibinfo {year} {2019})}\BibitemShut {NoStop}%
\bibitem [{\citenamefont {Sgrignuoli}\ \emph
  {et~al.}(2019{\natexlab{a}})\citenamefont {Sgrignuoli}, \citenamefont {Wang},
  \citenamefont {Pinheiro},\ and\ \citenamefont {Dal~Negro}}]{Sgrignuoli2019}%
  \BibitemOpen
  \bibfield  {author} {\bibinfo {author} {\bibfnamefont {F.}~\bibnamefont
  {Sgrignuoli}}, \bibinfo {author} {\bibfnamefont {R.}~\bibnamefont {Wang}},
  \bibinfo {author} {\bibfnamefont {F.~A.}\ \bibnamefont {Pinheiro}}, \ and\
  \bibinfo {author} {\bibfnamefont {L.}~\bibnamefont {Dal~Negro}},\ }\bibfield
  {title} {\emph {\bibinfo {title} {Localization of scattering resonances in
  aperiodic Vogel spirals},}\ }\href@noop {} {\bibfield  {journal} {\bibinfo
  {journal} {Phys. Rev. B}\ }\textbf {\bibinfo {volume} {99}},\ \bibinfo
  {pages} {104202} (\bibinfo {year} {2019}{\natexlab{a}})}\BibitemShut
  {NoStop}%
\bibitem [{\citenamefont {Sgrignuoli}\ \emph
  {et~al.}(2019{\natexlab{b}})\citenamefont {Sgrignuoli}, \citenamefont
  {R{\"o}ntgen}, \citenamefont {Morfonios}, \citenamefont {Schmelcher},\ and\
  \citenamefont {Dal~Negro}}]{Sgrignuoli2019compact}%
  \BibitemOpen
  \bibfield  {author} {\bibinfo {author} {\bibfnamefont {F.}~\bibnamefont
  {Sgrignuoli}}, \bibinfo {author} {\bibfnamefont {M.}~\bibnamefont
  {R{\"o}ntgen}}, \bibinfo {author} {\bibfnamefont {C.~V.}\ \bibnamefont
  {Morfonios}}, \bibinfo {author} {\bibfnamefont {P.}~\bibnamefont
  {Schmelcher}}, \ and\ \bibinfo {author} {\bibfnamefont {L.}~\bibnamefont
  {Dal~Negro}},\ }\bibfield  {title} {\emph {\bibinfo {title} {Compact
  localized states of open scattering media: a graph decomposition approach for
  an ab initio design},}\ }\href@noop {} {\bibfield  {journal} {\bibinfo
  {journal} {Opt. Lett.}\ }\textbf {\bibinfo {volume} {44}},\ \bibinfo {pages}
  {375} (\bibinfo {year} {2019}{\natexlab{b}})}\BibitemShut {NoStop}%
\bibitem [{\citenamefont {Skipetrov}\ and\ \citenamefont
  {Sokolov}(2014)}]{SkipetrovPRL}%
  \BibitemOpen
  \bibfield  {author} {\bibinfo {author} {\bibfnamefont {S.~E.}\ \bibnamefont
  {Skipetrov}}\ and\ \bibinfo {author} {\bibfnamefont {I.~M.}\ \bibnamefont
  {Sokolov}},\ }\bibfield  {title} {\emph {\bibinfo {title} {Absence of
  Anderson localization of light in a random ensemble of point scatterers},}\
  }\href@noop {} {\bibfield  {journal} {\bibinfo  {journal} {Phys. Rev. Lett.}\
  }\textbf {\bibinfo {volume} {112}},\ \bibinfo {pages} {023905} (\bibinfo
  {year} {2014})}\BibitemShut {NoStop}%
\bibitem [{\citenamefont {Caz{\'e}}\ \emph {et~al.}(2013)\citenamefont
  {Caz{\'e}}, \citenamefont {Pierrat},\ and\ \citenamefont {Carminati}}]{Caze}%
  \BibitemOpen
  \bibfield  {author} {\bibinfo {author} {\bibfnamefont {A.}~\bibnamefont
  {Caz{\'e}}}, \bibinfo {author} {\bibfnamefont {R.}~\bibnamefont {Pierrat}}, \
  and\ \bibinfo {author} {\bibfnamefont {R.}~\bibnamefont {Carminati}},\
  }\bibfield  {title} {\emph {\bibinfo {title} {Strong coupling to
  two-dimensional Anderson localized modes},}\ }\href@noop {} {\bibfield
  {journal} {\bibinfo  {journal} {Phys. Rev. Lett.}\ }\textbf {\bibinfo
  {volume} {111}},\ \bibinfo {pages} {053901} (\bibinfo {year}
  {2013})}\BibitemShut {NoStop}%
\bibitem [{\citenamefont {Bouchet}\ and\ \citenamefont
  {Carminati}(2019)}]{Bouchet}%
  \BibitemOpen
  \bibfield  {author} {\bibinfo {author} {\bibfnamefont {D.}~\bibnamefont
  {Bouchet}}\ and\ \bibinfo {author} {\bibfnamefont {R.}~\bibnamefont
  {Carminati}},\ }\bibfield  {title} {\emph {\bibinfo {title} {Quantum dipole
  emitters in structured environments: a scattering approach: tutorial},}\
  }\href@noop {} {\bibfield  {journal} {\bibinfo  {journal} {JOSA A}\ }\textbf
  {\bibinfo {volume} {36}},\ \bibinfo {pages} {186} (\bibinfo {year}
  {2019})}\BibitemShut {NoStop}%
\bibitem [{\citenamefont {Sgrignuoli}\ \emph {et~al.}(2015)\citenamefont
  {Sgrignuoli}, \citenamefont {Mazzamuto}, \citenamefont {Caselli},
  \citenamefont {Intonti}, \citenamefont {Cataliotti}, \citenamefont
  {Gurioli},\ and\ \citenamefont {Toninelli}}]{SgrignuoliACS}%
  \BibitemOpen
  \bibfield  {author} {\bibinfo {author} {\bibfnamefont {F.}~\bibnamefont
  {Sgrignuoli}}, \bibinfo {author} {\bibfnamefont {G.}~\bibnamefont
  {Mazzamuto}}, \bibinfo {author} {\bibfnamefont {N.}~\bibnamefont {Caselli}},
  \bibinfo {author} {\bibfnamefont {F.}~\bibnamefont {Intonti}}, \bibinfo
  {author} {\bibfnamefont {F.~S.}\ \bibnamefont {Cataliotti}}, \bibinfo
  {author} {\bibfnamefont {M.}~\bibnamefont {Gurioli}}, \ and\ \bibinfo
  {author} {\bibfnamefont {C.}~\bibnamefont {Toninelli}},\ }\bibfield  {title}
  {\emph {\bibinfo {title} {Necklace state hallmark in disordered 2D photonic
  systems},}\ }\href@noop {} {\bibfield  {journal} {\bibinfo  {journal} {ACS
  Photonics}\ }\textbf {\bibinfo {volume} {2}},\ \bibinfo {pages} {1636}
  (\bibinfo {year} {2015})}\BibitemShut {NoStop}%
\bibitem [{\citenamefont {Mirlin}(2000)}]{Mirlin}%
  \BibitemOpen
  \bibfield  {author} {\bibinfo {author} {\bibfnamefont {A.~D.}\ \bibnamefont
  {Mirlin}},\ }\bibfield  {title} {\emph {\bibinfo {title} {Statistics of
  energy levels and eigenfunctions in disordered systems},}\ }\href@noop {}
  {\bibfield  {journal} {\bibinfo  {journal} {Physics Reports}\ }\textbf
  {\bibinfo {volume} {326}},\ \bibinfo {pages} {259} (\bibinfo {year}
  {2000})}\BibitemShut {NoStop}%
\bibitem [{\citenamefont {Sgrignuoli}\ \emph {et~al.}(2020)\citenamefont
  {Sgrignuoli}, \citenamefont {Gorsky}, \citenamefont {Britton}, \citenamefont
  {Zhang}, \citenamefont {Riboli},\ and\ \citenamefont
  {Dal~Negro}}]{Sgrignuoli_MF}%
  \BibitemOpen
  \bibfield  {author} {\bibinfo {author} {\bibfnamefont {F.}~\bibnamefont
  {Sgrignuoli}}, \bibinfo {author} {\bibfnamefont {S.}~\bibnamefont {Gorsky}},
  \bibinfo {author} {\bibfnamefont {W.~A.}\ \bibnamefont {Britton}}, \bibinfo
  {author} {\bibfnamefont {R.}~\bibnamefont {Zhang}}, \bibinfo {author}
  {\bibfnamefont {F.}~\bibnamefont {Riboli}}, \ and\ \bibinfo {author}
  {\bibfnamefont {L.}~\bibnamefont {Dal~Negro}},\ }\bibfield  {title} {\emph
  {\bibinfo {title} {Multifractality of light in photonic arrays based on
  algebraic number theory},}\ }\href@noop {} {\bibfield  {journal} {\bibinfo
  {journal} {Commun. Phys.}\ }\textbf {\bibinfo {volume} {3}},\ \bibinfo
  {pages} {1} (\bibinfo {year} {2020})}\BibitemShut {NoStop}%
\bibitem [{\citenamefont {Ryu}\ \emph {et~al.}(1992)\citenamefont {Ryu},
  \citenamefont {Oh},\ and\ \citenamefont {Lee}}]{RyuPRB}%
  \BibitemOpen
  \bibfield  {author} {\bibinfo {author} {\bibfnamefont {C.}~\bibnamefont
  {Ryu}}, \bibinfo {author} {\bibfnamefont {G.}~\bibnamefont {Oh}}, \ and\
  \bibinfo {author} {\bibfnamefont {M.}~\bibnamefont {Lee}},\ }\bibfield
  {title} {\emph {\bibinfo {title} {Extended and critical wave functions in a
  Thue-Morse chain},}\ }\href@noop {} {\bibfield  {journal} {\bibinfo
  {journal} {Phys. Rev. B}\ }\textbf {\bibinfo {volume} {46}},\ \bibinfo
  {pages} {5162} (\bibinfo {year} {1992})}\BibitemShut {NoStop}%
\bibitem [{\citenamefont {Maci{\'a}}(1999)}]{macia1999}%
  \BibitemOpen
  \bibfield  {author} {\bibinfo {author} {\bibfnamefont {E.}~\bibnamefont
  {Maci{\'a}}},\ }\bibfield  {title} {\emph {\bibinfo {title} {Physical nature
  of critical modes in Fibonacci quasicrystals},}\ }\href@noop {} {\bibfield
  {journal} {\bibinfo  {journal} {Phys. Rev. B}\ }\textbf {\bibinfo {volume}
  {60}},\ \bibinfo {pages} {10032} (\bibinfo {year} {1999})}\BibitemShut
  {NoStop}%
\bibitem [{\citenamefont {Escalante}\ and\ \citenamefont
  {Skipetrov}(2018)}]{Escalante}%
  \BibitemOpen
  \bibfield  {author} {\bibinfo {author} {\bibfnamefont {J.~M.}\ \bibnamefont
  {Escalante}}\ and\ \bibinfo {author} {\bibfnamefont {S.~E.}\ \bibnamefont
  {Skipetrov}},\ }\bibfield  {title} {\emph {\bibinfo {title} {Level spacing
  statistics for light in two-dimensional disordered photonic crystals},}\
  }\href@noop {} {\bibfield  {journal} {\bibinfo  {journal} {Sci. Rep.}\
  }\textbf {\bibinfo {volume} {8}},\ \bibinfo {pages} {11569} (\bibinfo {year}
  {2018})}\BibitemShut {NoStop}%
\bibitem [{\citenamefont {Skipetrov}\ and\ \citenamefont
  {Sokolov}(2015)}]{Skipetrov2015}%
  \BibitemOpen
  \bibfield  {author} {\bibinfo {author} {\bibfnamefont {S.~E.}\ \bibnamefont
  {Skipetrov}}\ and\ \bibinfo {author} {\bibfnamefont {I.~M.}\ \bibnamefont
  {Sokolov}},\ }\bibfield  {title} {\emph {\bibinfo {title}
  {Magnetic-field-driven localization of light in a cold-atom gas},}\
  }\href@noop {} {\bibfield  {journal} {\bibinfo  {journal} {Phys. Rev. Lett.}\
  }\textbf {\bibinfo {volume} {114}},\ \bibinfo {pages} {053902} (\bibinfo
  {year} {2015})}\BibitemShut {NoStop}%
\bibitem [{\citenamefont {Mondal}\ \emph {et~al.}(2019)\citenamefont {Mondal},
  \citenamefont {Kumar}, \citenamefont {Kamp},\ and\ \citenamefont
  {Mujumdar}}]{Mondal}%
  \BibitemOpen
  \bibfield  {author} {\bibinfo {author} {\bibfnamefont {S.}~\bibnamefont
  {Mondal}}, \bibinfo {author} {\bibfnamefont {R.}~\bibnamefont {Kumar}},
  \bibinfo {author} {\bibfnamefont {M.}~\bibnamefont {Kamp}}, \ and\ \bibinfo
  {author} {\bibfnamefont {S.}~\bibnamefont {Mujumdar}},\ }\bibfield  {title}
  {\emph {\bibinfo {title} {Optical Thouless conductance and level-spacing
  statistics in two-dimensional Anderson localizing systems},}\ }\href@noop {}
  {\bibfield  {journal} {\bibinfo  {journal} {Phys. Rev. B}\ }\textbf {\bibinfo
  {volume} {100}},\ \bibinfo {pages} {060201} (\bibinfo {year}
  {2019})}\BibitemShut {NoStop}%
\bibitem [{\citenamefont {Wang}\ \emph {et~al.}(2018)\citenamefont {Wang},
  \citenamefont {Pinheiro},\ and\ \citenamefont {Dal~Negro}}]{Wang_Prime}%
  \BibitemOpen
  \bibfield  {author} {\bibinfo {author} {\bibfnamefont {R.}~\bibnamefont
  {Wang}}, \bibinfo {author} {\bibfnamefont {F.~A.}\ \bibnamefont {Pinheiro}},
  \ and\ \bibinfo {author} {\bibfnamefont {L.}~\bibnamefont {Dal~Negro}},\
  }\bibfield  {title} {\emph {\bibinfo {title} {Spectral statistics and
  scattering resonances of complex primes arrays},}\ }\href@noop {} {\bibfield
  {journal} {\bibinfo  {journal} {Phys. Rev. B}\ }\textbf {\bibinfo {volume}
  {97}},\ \bibinfo {pages} {024202} (\bibinfo {year} {2018})}\BibitemShut
  {NoStop}%
\bibitem [{\citenamefont {Haake}(1991)}]{Haake}%
  \BibitemOpen
  \bibfield  {author} {\bibinfo {author} {\bibfnamefont {F.}~\bibnamefont
  {Haake}},\ }\bibfield  {title} {\emph {\bibinfo {title} {Quantum signatures
  of chaos},}\ }in\ \href@noop {} {\emph {\bibinfo {booktitle} {Quantum
  Coherence in Mesoscopic Systems}}}\ (\bibinfo  {publisher} {Springer},\
  \bibinfo {year} {1991})\ pp.\ \bibinfo {pages} {583--595}\BibitemShut
  {NoStop}%
\bibitem [{\citenamefont {Skipetrov}(2016)}]{SkipetrovPRB}%
  \BibitemOpen
  \bibfield  {author} {\bibinfo {author} {\bibfnamefont {S.}~\bibnamefont
  {Skipetrov}},\ }\bibfield  {title} {\emph {\bibinfo {title} {Finite-size
  scaling analysis of localization transition for scalar waves in a
  three-dimensional ensemble of resonant point scatterers},}\ }\href@noop {}
  {\bibfield  {journal} {\bibinfo  {journal} {Phys. Rev. B}\ }\textbf {\bibinfo
  {volume} {94}},\ \bibinfo {pages} {064202} (\bibinfo {year}
  {2016})}\BibitemShut {NoStop}%
\bibitem [{\citenamefont {Sheng}(2006)}]{Sheng}%
  \BibitemOpen
  \bibfield  {author} {\bibinfo {author} {\bibfnamefont {P.}~\bibnamefont
  {Sheng}},\ }\href@noop {} {\emph {\bibinfo {title} {Introduction to wave
  scattering, localization and mesoscopic phenomena}}},\ Vol.~\bibinfo {volume}
  {88}\ (\bibinfo  {publisher} {Springer Science \& Business Media},\ \bibinfo
  {year} {2006})\BibitemShut {NoStop}%
\bibitem [{\citenamefont {Gupta}\ and\ \citenamefont {Ye}(2003)}]{Gupta}%
  \BibitemOpen
  \bibfield  {author} {\bibinfo {author} {\bibfnamefont {B.~C.}\ \bibnamefont
  {Gupta}}\ and\ \bibinfo {author} {\bibfnamefont {Z.}~\bibnamefont {Ye}},\
  }\bibfield  {title} {\emph {\bibinfo {title} {Localization of classical waves
  in two-dimensional random media: A comparison between the analytic theory and
  exact numerical simulation},}\ }\href@noop {} {\bibfield  {journal} {\bibinfo
   {journal} {Phys. Rev E}\ }\textbf {\bibinfo {volume} {67}},\ \bibinfo
  {pages} {036606} (\bibinfo {year} {2003})}\BibitemShut {NoStop}%
\bibitem [{\citenamefont {Skipetrov}\ \emph {et~al.}(2020)\citenamefont
  {Skipetrov} \emph {et~al.}}]{Skipetrov2020finite}%
  \BibitemOpen
  \bibfield  {author} {\bibinfo {author} {\bibfnamefont {S.~E.}\ \bibnamefont
  {Skipetrov}} \emph {et~al.},\ }\bibfield  {title} {\emph {\bibinfo {title}
  {Finite-size scaling of the density of states inside band gaps of ideal and
  disordered photonic crystals},}\ }\href@noop {} {\bibfield  {journal}
  {\bibinfo  {journal} {Eur. Phys. J. B}\ }\textbf {\bibinfo {volume} {93}},\
  \bibinfo {pages} {1} (\bibinfo {year} {2020})}\BibitemShut {NoStop}%
\bibitem [{\citenamefont {Joannopoulos}\ \emph {et~al.}(1995)\citenamefont
  {Joannopoulos}, \citenamefont {Johnson}, \citenamefont {Winn},\ and\
  \citenamefont {Meade}}]{Joannopoulos}%
  \BibitemOpen
  \bibfield  {author} {\bibinfo {author} {\bibfnamefont {J.~D.}\ \bibnamefont
  {Joannopoulos}}, \bibinfo {author} {\bibfnamefont {S.~G.}\ \bibnamefont
  {Johnson}}, \bibinfo {author} {\bibfnamefont {J.~N.}\ \bibnamefont {Winn}}, \
  and\ \bibinfo {author} {\bibfnamefont {R.~D.}\ \bibnamefont {Meade}},\
  }\href@noop {} {\emph {\bibinfo {title} {Photonic crystals: Molding the flow
  of light}}}\ (\bibinfo  {publisher} {Princeton University Press Princeton,
  NJ},\ \bibinfo {year} {1995})\BibitemShut {NoStop}%
\bibitem [{\citenamefont {Hasan}\ \emph {et~al.}(2018)\citenamefont {Hasan},
  \citenamefont {Mosk}, \citenamefont {Vos},\ and\ \citenamefont
  {Lagendijk}}]{hasan2018finite}%
  \BibitemOpen
  \bibfield  {author} {\bibinfo {author} {\bibfnamefont {S.~B.}\ \bibnamefont
  {Hasan}}, \bibinfo {author} {\bibfnamefont {A.~P.}\ \bibnamefont {Mosk}},
  \bibinfo {author} {\bibfnamefont {W.~L.}\ \bibnamefont {Vos}}, \ and\
  \bibinfo {author} {\bibfnamefont {A.}~\bibnamefont {Lagendijk}},\ }\bibfield
  {title} {\emph {\bibinfo {title} {Finite-size scaling of the density of
  states in photonic band gap crystals},}\ }\href@noop {} {\bibfield  {journal}
  {\bibinfo  {journal} {Phys. Rev. Lett.}\ }\textbf {\bibinfo {volume} {120}},\
  \bibinfo {pages} {237402} (\bibinfo {year} {2018})}\BibitemShut {NoStop}%
\end{thebibliography}
\end{document}